\documentclass[twocolumn,amsmath,amssymb,floatfix,nofootinbib]{revtex4-1}

\usepackage{graphicx}
\usepackage{dcolumn}
\usepackage{bm}
\usepackage{latexsym}
\usepackage{slashed}
\usepackage[colorlinks=true,linkcolor=blue,citecolor=blue]{hyperref}
\usepackage[dvipsnames]{xcolor}
\usepackage[normalem]{ulem} 

\begin{document}

\title{Non-resonant particle acceleration in strong turbulence: comparison to kinetic and MHD simulations}

\author{Virginia Bresci} 
\affiliation{Institut d'Astrophysique de Paris,\\
CNRS -- Sorbonne Universit\'e, \\
98 bis boulevard Arago, F-75014 Paris, France}
\affiliation{CEA, DAM, DIF, F-91297 Arpajon, France}
\author{Martin Lemoine} 
\affiliation{Institut d'Astrophysique de Paris,\\
CNRS -- Sorbonne Universit\'e, \\
98 bis boulevard Arago, F-75014 Paris, France}
\author{Laurent Gremillet} 
\affiliation{CEA, DAM, DIF, F-91297 Arpajon, France}
\affiliation{Universit\'e Paris-Saclay, CEA, LMCE, 91680 Bruy\`eres-le-Ch\^atel, France}
\author{Luca Comisso} 
\affiliation{Department of Astronomy and Columbia Astrophysics Laboratory, Columbia University, New York, NY 10027, USA}
\author{Lorenzo Sironi} 
\affiliation{Department of Astronomy and Columbia Astrophysics Laboratory, Columbia University, New York, NY 10027, USA}
\author{Camilia Demidem} 
\affiliation{Nordita,\\ KTH Royal Institute of Technology and Stockholm University,\\ Hannes Alfv\'{e}ns v\"{a}g 12, SE-106 91 Stockholm, Sweden}
\affiliation{JILA,\\ University of Colorado and National Institute of Standards and Technology,\\ 440 UCB, Boulder, CO 80309-0440, USA}

\date{\today}

\begin{abstract} 
Collisionless, magnetized turbulence offers a promising framework for the generation of non-thermal high-energy particles in various astrophysical sites. Yet, the detailed mechanism that governs particle acceleration has remained subject to debate. By means of 2D and 3D PIC, as well as 3D (incompressible) magnetohydrodynamic (MHD) simulations, we test here a recent model of non-resonant particle acceleration in strongly magnetized turbulence~\cite{2021PhRvD.104f3020L}, which ascribes the energization of particles to their continuous interaction with the random velocity flow of the turbulence, in the spirit of the original Fermi model. To do so, we compare, for a large number of particles that were tracked in the simulations, the predicted and the observed histories of particles momenta. The predicted history is that derived from the model, after extracting from the simulations, at each point along the particle trajectory, the three force terms that control acceleration: the acceleration of the field line velocity projected along the field line direction, its shear projected along the same direction, and its transverse compressive part. Overall, we find a clear correlation between the model predictions and the numerical experiments, indicating that this non-resonant model can successfully account for the bulk of particle energization through Fermi-type processes in strongly magnetized turbulence. We also observe that the parallel shear contribution tends to dominate the physics of energization in the PIC simulations, while in the MHD incompressible simulation, both the parallel shear and the transverse compressive term provide about equal contributions. 
\end{abstract}

\maketitle

\section{Introduction}
Stochastic particle acceleration in magnetized turbulence has emerged as a key process in high-energy astrophysics, as it is likely at play in a wide variety of sources and in a diverse set of physical conditions, from the solar atmosphere~\cite{2004ApJ...610..550P,2004MNRAS.354..870S,2012ApJ...754..103B} up to more exotic objects, {\it e.g.} blazars, gamma-ray bursts and other relativistic outflows~\cite{1996ApJ...461L..37B,2011ApJ...739...66T,2016JPlPh..82d6301L,2016PhRvD..94b3005A,2017ApJ...846L..28X,18Asano,2019ApJ...872...10X}. In phenomenological applications, particle acceleration is commonly characterized by a diffusion coefficient in momentum space, whose magnitude and scaling control the evolution of the particle distribution function in a Fokker-Planck description. Most often, this diffusion coefficient is calculated in the framework of quasilinear theory, which ascribes particle energization to resonant interactions with linear eigenmodes of the plasma, {\it e.g.}~\cite{2002cra..book.....S,2008ApJ...681.1725S}. Quasilinear theory, however, is a perturbative description restricted to the regime of small-amplitude turbulence, {\it i.e.}  $\delta B\ll B$, with $\delta B$ the turbulent rms magnetic field fluctuation on the integral scale $\ell_{\rm c}$ of the turbulence cascade, and $B$ the mean field strength. To the contrary, the turbulence can be regarded as of large-amplitude in most astrophysical settings. Furthermore, stochastic acceleration is expected to be rather slow in the small-amplitude limit, since the acceleration timescale, {\it i.e.} the time needed on average to double a particle energy, scales in proportion to $\left(\delta B/B\right)^{-2}$.

The regime of strong turbulence is thus of broader applicability and interest, if not for practical purposes. Recently, kinetic particle-in-cell (PIC) simulations have offered {\it ab initio} numerical probes of that regime, in the relativistic limit where the Alfv\'en velocity $v_{\rm A}\sim c$~\cite{17Zhdankin,18Comisso,2018ApJ...867L..18Z,2018MNRAS.474.2514Z,2019ApJ...886..122C,20Comisso,2020ApJ...893L...7W,2019PhRvL.122e5101Z,2020arXiv201203043N,21Comisso,2021ApJ...922..172Z}. Quite interestingly, those experiments have measured a momentum diffusion coefficient $D_{pp}\sim 0.1 \,\langle \delta u^2\rangle \,\,p^2 c /\ell_{\rm c}$, here written in terms of the variance of four-velocity fluctuations $\langle \delta u^2\rangle$ and particle momentum $p$, whose $p^2$ dependence is suggestive of a non-resonant form of acceleration.

Non-resonant acceleration can be described in various ways. In the original Fermi picture~\cite{49Fermi}, acceleration proceeds through repeated encounters with discrete, point-like moving magnetized structures. The particle then draws energy from the motional electric field carried by those highly conducting plasma elements; accordingly, in the reference frame of the scattering center, the interaction is assumed to be purely elastic and described as pitch-angle scattering. If the mean free path to interaction $t_{\rm mfp}$ does not depend on energy, and if the particle distribution function is assumed isotropic before each interaction, then the energization process can be indeed described as a random walk in momentum space with diffusion coefficient $D_{pp} \propto p^2/t_{\rm mfp}$~\cite{49Fermi,1954ApJ...119....1F,2005PPCF...47B.667D}.

Generalizing this picture to a turbulent flow, the acceleration of a particle can now be related in general terms to the time variation of the velocity flow that the particle encounters along its journey, {\it i.e.} to its interaction with the time-dependent sheared and compressive parts of the drift velocity field $\boldsymbol{v_E}=\boldsymbol{E}\times\boldsymbol{B}/B^2$ (in units of $c$), which characterizes the velocity of magnetic field lines in the magnetohydrodynamics (MHD) approximation~\cite{1983ICRC....9..313B,1988SvAL...14..255P,1990A&A...236..519D,2003ApJ...595..195W,2004ApJ...603...23C,2010ApJ...713..475J,2019PhRvD..99h3006L,2021PhRvD.104f3020L}. More specifically, for particles whose gyroradius $r_{\rm g}$ is much smaller than the integral scale $\ell_{\rm c}$, non-resonant acceleration derives from three main force terms expressed in terms of $\boldsymbol{u_E}$, the four-velocity generalizing $\boldsymbol{v_E}$: an inertial contribution characterizing the time-dependence of $\boldsymbol{u_E}$, its shear as projected along the mean magnetic field direction, and its compression in the plane transverse to the mean magnetic field~\cite{2021PhRvD.104f3020L}. Those force terms and their overall combination will be detailed further below.

The objective of the present paper is to test this non-resonant acceleration model through a direct comparison to numerical simulations of strong turbulence. To do so, we rely on dedicated large-scale PIC simulations of forced turbulence in 2D and 3D, as well as of decaying turbulence in 2D; we also make use of an incompressible MHD simulation, borrowed from the Johns Hopkins University turbulence database. We frame the discussion as follows. In Sec.~\ref{sec:model}, we detail the theoretical model and explain the metric that we use to test it on numerical simulations. In Sec.~\ref{sec:num}, we then present the numerical simulations and carry out our test. This comparison, which proves quite satisfactory, is summarized and discussed in Sec.~\ref{sec:conc}. Finally, we present some similar tests of our model against simulations of test-particle acceleration in synthetic wave turbulence in App.~\ref{app:synthetic}, which are known to be well described by quasilinear theory, even at relatively large turbulence amplitude $\delta B/B \sim 1$.

\section{Theoretical model}\label{sec:model}
\subsection{Acceleration models}\label{sec:th-prelim}

Let us first emphasize some differences between the more conventional approach based on quasilinear theory and the non-resonant acceleration scenario that we study here; see also \cite{2021PhRvD.104f3020L} for a detailed discussion of those issues. In its simplest formulation, quasilinear theory describes the turbulence as a linear superposition of eigenmodes (fast and slow MHD magnetosonic waves, and Alfv\'en waves in the MHD approximation) of infinite spatial extent and well defined frequency~\cite{2002cra..book.....S}. Nonlinear extensions include resonance-broadening effects or stochastic corrections to the particle orbits, see~\cite{2020PhRvD.102b3003D} for a recent general discussion, and references therein. In the framework of modern, anisotropic MHD turbulence theories~\cite{95GS,1997ApJ...485..680G}, resonance broadening effects associated with wave damping are actually necessary to restore a part of wave-particle interactions, which would otherwise disappear at small $r_{\rm g}/\ell_{\rm c}$~\cite{2000PhRvL..85.4656C}, with the exception of fast MHD modes, which preserve resonant interactions~\cite{2002PhRvL..89B1102Y}.
Such resonant interactions would lead to a scaling $D_{pp}\propto p^{q}$, with $q$ the (absolute value of the) index of the one-dimensional power spectrum of magnetic fluctuations contained in fast modes. This disagrees with the scaling observed in recent PIC simulations, if $q<2$ as expected and observed~\cite{03Cho,16Takamoto}\footnote{Ref.~\cite{17Takamoto} reports a spectrum characterized by $q\simeq 2$ for fast modes in \emph{isothermal} compressible, relativistic turbulence; however, this isothermal configuration, which is obtained by adding an ad-hoc cooling term to the plasma evolution, does not match the conditions of the PIC simulations of relativistic turbulence which have reported $D_{pp}\propto p^2$.}.

By contrast, resonance-broadened interactions yield $D_{pp}\propto p^2$ up to logarithmic corrections~\cite{2000PhRvL..85.4656C,2020PhRvD.102b3003D}. This scaling emerges because, once the resonance is broadened, all scales above the gyroradius give equal contributions to the diffusion coefficient. In the context of strong turbulence, this contradicts the quasilinear assumption, because the wave-particle interaction timescale ($\sim \ell_{\rm c}/c$) becomes larger than the timescale over which the magnetic field strength and direction have changed by the order of unity. Moreover, in such turbulence, the eigenmodes that contribute on all scales up to $\ell_{\rm c}$ are aligned with respect to different mean field directions, since this orientation changes with scale~\cite{2000ApJ...539..273C}. The polarization of those misaligned eddies, as seen by the particle, thus does not correspond to that of linear eigenmodes, further quelling any resonance. From its point of view, meaning as seen on a scale $r_{\rm g}$, the particle rather experiences those various modes as random velocity structures distributed on all scales up to $\ell_{\rm c}$, and it interacts with them in a non-resonant way.

In that sense, non-resonant acceleration as we discuss it is not antagonistic to the presence of waves or wavepackets in a turbulent bath; it assumes, however, that sharp wave-particle resonances are absent, or at least sub-dominant. The results that we present further below will support that point of view.
 
Although the non-resonant acceleration process that we test here can be regarded as a generalization of the original Fermi scenario to a continuous turbulent flow, there are noteworthy differences. In particular, the particle does not gain energy depending on whether its interaction with the velocity structure is head-on or tail-on in the present case.
It rather gains, or loses energy, depending on the sign of the variation of the velocity flow while inside the structure~\cite{2019PhRvD..99h3006L,2021PhRvD.104f3020L}. For instance, in a region that undergoes compression, energy gain is positive, while it is negative if the flow undergoes dilation, as discussed further below. That difference can be traced back to the underlying assumptions: the original model of E. Fermi depicts discrete, point-like interactions in a fixed laboratory frame, while our non-resonant model rather discusses how particles interact with modes of extent $\gtrsim r_{\rm g}$ in a comoving frame. The two pictures do not contradict each other, however. To see this, consider a moving magnetic mirror (type-A interaction in the Fermi model): in the comoving frame, the particle will experience in (comoving) time a compression or a dilation depending on whether the mirror is moving towards (head-on) or away from (tail-on) the particle, leading respectively to energy gain or loss. 

\subsection{Non-resonant acceleration}\label{sec:nr-acc}

Non-resonant acceleration is best described by making use of the formalism developed in \cite{2019PhRvD..99h3006L,2021PhRvD.104f3020L}, which follows the momentum of the particle in a sequence of frames in which the electric field vanishes. One assumption of that model is  that this reference frame  exists, meaning that the Lorentz invariant quantity $\boldsymbol{E^2}-\boldsymbol{B^2}$ is everywhere negative (or at least, over most of space). This is in particular satisfied in the ideal MHD approximation, in which case $\boldsymbol{E}\,=\,-\boldsymbol{v_{\rm p}}\times\boldsymbol{B}$ (henceforth all velocities are written in units of $c$), with $\boldsymbol{v_{\rm p}}$ the plasma bulk velocity. We assume here that this MHD regime applies. In the following, the frame in which $\boldsymbol{E}$ vanishes is denoted $\mathcal R_{\slashed E}$; with respect to the laboratory frame, it moves with the velocity $\boldsymbol{v_E}$ defined earlier. Quantities expressed in $\mathcal R_{\slashed E}$ will be annotated with a prime; while this distinction does not matter in a non-relativistic setting, it matters here, because we will compare our results to PIC simulations of relativistic turbulence. We emphasize, however, that the present treatment is general, and that it can be applied equally well to the sub-relativistic or to the relativistic regime. 

The advantage of this approach is that it allows one to substitute the electric field for the gradients of $\boldsymbol{v_E}$ in the expression of forces acting on the particles. Energy gains and losses are thus directly connected to the statistics of the velocity structures. At this stage, this can be seen as plain rewriting, no information has been lost. For practical purposes, though, this model can be simplified further by noting that turbulent modes of wavelengths significantly smaller than the gyroradius of the particle contribute little, if at all, to the process of energization. Retaining only the contribution of modes of scales $\gtrsim r_{\rm g}$, one can approximate the gradients as their average over a gyrating orbit of the particle around the field line. Note that we do not follow unperturbed trajectories around a mean field line, as in quasilinear theory, but the exact trajectory around the perturbed field line.

In that approximation, the energization of the particle can be written in terms of three force terms only, which are proportional to the quantities $\Theta_\parallel$, $\Theta_\perp$ and $\boldsymbol{a_{\rm E}}\cdot\boldsymbol{b}$. These have the following definitions and interpretations:
\begin{equation}
\Theta_\parallel \,=\,b^\alpha b^\beta \partial_\alpha {u_E}_\beta\,
\end{equation}
denotes the projection of the shear of the (field line) four-velocity field $\boldsymbol{u_E}=\boldsymbol{v_E}/\sqrt{1-v_E^2}$ along the direction of the mean field direction $\boldsymbol{b}$ ($\boldsymbol{b}$ a unit vector). Here $\alpha,\,\beta\,\in\,\left\{0,\ldots,3\right\}$ are spacetime indices, $b^\alpha=\left\{0,\boldsymbol{b}\right\}$ and ${u_E}^\alpha=\left\{\gamma_E,\boldsymbol{u_E}\right\}$.

The mean-field direction is interpreted as the sum of the coherent magnetic field $\boldsymbol{B_0}$ and of all modes on scales larger than the gyroradius, {\it i.e.}, $\boldsymbol{b} = \boldsymbol{\overline B}/\overline B$ with
\begin{equation}
\boldsymbol{\overline B}\left(\boldsymbol{x},\,t;\,\boldsymbol{p}\right)\,=\,\boldsymbol{B_0} + \boldsymbol{\delta B_l}\left(\boldsymbol{x},\,t;\,\boldsymbol{p}\right)\,,
\label{eq:meanB}
\end{equation}
where $\boldsymbol{\delta B_l}$ represents the perturbation coarse-grained on scale $l=r_{\rm g}$. It should be understood as a fluctuation seen at that position, after filtering out scales smaller than $l$. Consequently, the mean field is here  a function of both particle position and momentum.

The force term associated with $\Theta_\parallel$ leads to energy gain if negative, to energy loss if positive. It can be related to the projection on $\boldsymbol{E}$ of the drift velocity (in a guiding center formulation) due to the curvature of the magnetic field line~\cite{2021PhRvD.104f3020L}. This term thus describes acceleration through the curvature term, or in the phrasing of the original Fermi mechanism, a Fermi type-B interaction which energizes the particle as in a slingshot.

The perpendicular shear term $\Theta_\perp$ is defined as
\begin{equation}
\Theta_\perp \,=\,\left(\eta^{\alpha\beta}  - 
b^\alpha b^\beta\right) \partial_\alpha {u_E}_\beta\,,
\end{equation}
with $\eta^{\alpha\beta}$ the Minkowski metric. It thus corresponds to the shear of the field line velocity field in the plane transverse to the magnetic field line. This term leads to energy gain or loss in much the same way as a particle gains or loses energy through the compression or expansion of a collisional plasma. Here, the particle is tied to the mean field through its orbit, thus transverse compression energizes the particle, while transverse expansion draws energy from it. This term also depicts the effect of a magnetic mirror, or Fermi type-A interaction, and as such it can be seen as a form of betatron acceleration~\cite{2021PhRvD.104f3020L}.

Finally, the acceleration term $\boldsymbol{a_E}\cdot\boldsymbol{b}$ depicts the influence of the effective gravity associated with the non-inertial nature of $\mathcal R_{\slashed E}$ as the particle travels along the field line. More explicitly, $\boldsymbol{a_E}$ is the (Lagrangian) three-acceleration of $\boldsymbol{u_E}$, {\it viz.}
\begin{equation}
\boldsymbol{a_E}\,=\,{u_E}^\beta\partial_\beta {u_E}^\alpha\,.
\end{equation}

With these definitions, the theoretical acceleration rate of a particle of mass $m$, as expressed in $\mathcal R_{\slashed E}$, can be written as
\begin{align}
\frac{1}{c}\left.\frac{{\rm d}\gamma'}{{\rm d}\tau}\right\vert_{\rm th.}&\,=\,-\gamma'u_\parallel'\,\boldsymbol{a_E}\cdot\boldsymbol{b} - {u_\parallel'}^2\,\Theta_\parallel- \frac{1}{2}{u_\perp'}^2\Theta_\perp\,,
\label{eq:Rco-evol}
\end{align}
with: $\gamma' = \epsilon'/mc^2$ (resp. $\epsilon'$) represents the Lorentz factor (resp. the energy) of the particle in  $\mathcal R_{\slashed E}$; $u_\parallel' = \boldsymbol{p'}\cdot \boldsymbol{b}/mc$ (resp $\boldsymbol{p'}$) its four-velocity projected along the mean magnetic field, in units of $c$ (resp. its three-momentum); correspondingly, $u_\perp'=\sqrt{{u'}^2-{u_\parallel'}^2}$ denotes the perpendicular component, with $u'=p'/mc$. Finally, ${\rm d}\tau\,=\,{\rm d}t'/\gamma'$ represents an element of proper time.

\subsection{Comparison to simulations}

We test the above non-resonant acceleration model against numerical simulations of particle acceleration in turbulence in the following way. For a given particle in a given simulation, we measure the history of its Lorentz factor, which we write $\gamma'_{\rm obs}(t)$, as a function of $t$  the time measured in the simulation frame.
In parallel, we reconstruct a theoretical history $\gamma'_{\rm th}(t)$ defined as follows in terms of the quantities that appear in Eq.~(\ref{eq:Rco-evol}):
\begin{equation}
\gamma'_{\rm th}(t)\,=\,\gamma'_{\rm obs}(t_0) + \int_{\tau(t_0)}^{\tau(t)}{\rm d}\tau\,\left.\frac{{\rm d}\gamma'}{{\rm d}\tau}\right\vert_{\rm th.}\,,
\label{eq:gamma-th}
\end{equation}
where $t_0$ represents some initial time. To do so, we extract from the numerical simulation, at each point of the trajectory, the various quantities that enter this equation, namely ${u_E}^\beta$ and its gradients $\partial_\alpha {u_E}^\beta$. We then reconstruct the fields $\boldsymbol{a_E}\cdot\boldsymbol{b}$, $\Theta_\parallel$ and $\Theta_\perp$ and use them to predict, at each time step of the trajectory, the change in $\gamma'$.

This reconstruction is affected by several effects. For one, the above model is an approximation obtained in the limit $r_{\rm g}\ll \ell_{\rm c}$, where $\ell_{\rm c}$ denotes the coherence scale of the turbulent power spectrum, {\it i.e.} the length scale on which most of the turbulent power lies. We therefore expect some effects of order $r_{\rm g}/\ell_{\rm c}$ to alter the model predictions. Because numerical simulations are restricted in their dynamic range (an effective rigidity $\rho = 2\pi r_{\rm g}/\ell_{\rm c} \sim 0.03 - 0.1$ is close to what that can be currently achieved at best), such effects can be significant, especially in regions of low magnetic field strength, as $r_{\rm g}$ can then take large values relative to its average at a given energy. Likewise, particles can experience substantial acceleration over a period of time, which also leads to an increase in $r_{\rm g}$, and hence to a loss of accuracy of the model predictions. 

Furthermore, the model assumes that the fields $\boldsymbol{u_E}$, $\boldsymbol{B'}$ and their gradients are coarse-grained quantities, meaning that sub-Larmor scales have been filtered out. Such a procedure is too costly to be put in place in PIC simulations, as the Larmor scale changes from particle to particle, and even from time step to time step, since the energy of a particle is not constant. We thus use the actual fields and gradients, as measured in the simulation on the scale of the numerical grid, and discard any filtering. This introduces high-frequency noise in the reconstruction of $\gamma'_{\rm th}$, associated with small-scale effects. To test how this may affect our comparison, we have also performed a reconstruction of the trajectory including time filtering, which allows us to smooth the time profile of the history of $\gamma'_{\rm obs}(t)$ on time scales $\simeq r_{\rm g}/c$. More explicitly, we smooth the fields $\boldsymbol{u_E}$, $\boldsymbol{B'}$ and their gradients that the particle encounters on its trajectory before performing the reconstruction. While the reconstructed trajectory differs from that obtained in the absence of this time filtering, the overall result remains similar.

Finally, the comparison with the model is made further complicated by non-ideal MHD effects. Kinetic simulations have demonstrated that particles of the thermal pool initially gain energy through non-ideal parallel electric field components in reconnection layers, then start to probe the large-scale turbulence once their gyroradius exceeds the typical scale of those layers~\cite{18Comisso,2019ApJ...886..122C}. The contribution of non-ideal parallel electric fields is observed to weaken with increasing particle energy, in general agreement with the idea that on large scales, the physics tends toward the ideal MHD regime, as assumed by the non-resonant acceleration model.
Those non-ideal electric fields, however, can affect the energy gain process on the simulation scales, and thus perturb the comparison. 
For this reason, on the one hand, we have evaluated in kinetic simulations the fraction of points along the particle trajectories where the ideal MHD condition is violated ($E >B$), which turns out not to exceed $0.05\,\%$ in 3D geometry. On the other hand, we have compared the model with the dynamics of test particles propagated in the fields extracted from a 3D ideal MHD simulation.

From the above considerations, we do not expect an exact match between $\gamma'_{\rm obs}$ and $\gamma'_{\rm th}$. To test the model, we thus calculate, for each particle $i$, a Pearson correlation coefficient $r_i$,
\begin{equation}
r_i\,\equiv\, \frac{{\rm cov}\left[\gamma'_{\rm obs}(i);\,\gamma'_{\rm th}(i)\right]}{{\rm cov}\left[\gamma'_{\rm obs}(i);\,\gamma'_{\rm obs}(i)\right]^{1/2}\,{\rm cov}\left[\gamma'_{\rm th}(i);\,\gamma'_{\rm th}(i)\right]^{1/2}}\,,
\label{eq:Pearson-c}
\end{equation}
where ${\rm cov}\left[A(i);\,B(i)\right]$ represents the covariance of the histories of the quantities $A$ and $B$ over the trajectory of particle $i$. 

We then collect, for a large number of particles, the sample of correlation coefficients and establish a probability density. We thus seek to see what fraction of those correlation coefficients lies within the vicinity of $+1$, that value denoting perfect correlation, hence perfect reconstruction.

As will be detailed in the following, sub-Larmor effects (or non-ideal electric fields on small length scales) can lead to a sharp departure in the history of $\gamma'_{\rm obs}$, and to a different departure in $\gamma'_{\rm th}$. This departure is sharp, because small-scale effects are associated with short timescales. For some particles, the two histories can thus reveal strong correlation before and after this sudden event, but the global trajectory itself will show a lesser degree of correlation. To bypass such effects, we have performed several reconstructions, which differ in the duration over which the trajectories are examined. In the following, we present reconstructions for intermediate timescales and for the entire duration of the simulation. The duration of the interval over which we follow the trajectories is written $\Delta t$. Additionally, we perform the correlation test for a sample of particles for which the energy varies by a significant factor, in order to test Eq.~(\ref{eq:Rco-evol}), treating on an equal footing energy gains and losses. To select the particles, we adopt a threshold $g_{\rm min}$ and consider those trajectories, or chunks of trajectories, that satisfy $\Delta\gamma'/\gamma'\geq g_{\rm min}$ with $\Delta\gamma'={\rm max}\left(\gamma'\right)-{\rm min}\left(\gamma'\right)$ over the interval of duration $\Delta t$. 

In the following Section, we describe the numerical simulations, the results of the reconstruction, including some details specific to each. In App.~\ref{app:synthetic}, we conduct a similar experiment on simulations that follow test particles in a synthetic turbulence, meaning a turbulence that is constructed from a sum of non-interacting linear eigenmodes (Alfv\'en, fast or slow magnetosonic modes) of the plasma, following the study of Ref.~\cite{2020PhRvD.102b3003D}. The interest of that experiment is that the physics of particle acceleration in such turbulence is relatively well understood, as it follows the predictions of quasilinear theory, and that part of it (transit-time damping acceleration related to magnetic mirroring effects) can be captured by the above model. It can therefore be used to gauge the amount of information contained in the probability density of correlation coefficients that we reconstruct.

\section{Numerical experiments}\label{sec:num}

In this section, we report on the comparison between the histories predicted by the model and those observed in several numerical experiments: (i) a 2D decaying turbulence $10\,000^2$ PIC simulation; (ii) a 2D forced turbulence $10\,000^2$ PIC simulation; (iii) a 3D forced turbulence $1\,080^3$ PIC simulation; and (iv) a 3D forced turbulence $1\,024^3$ MHD simulation.

All PIC simulations assume a pair plasma composition. They have been conducted using the finite-difference time-domain, relativistic PIC \textsc{calder} code~\cite{Lefebvre_2003}, to which a turbulence stirring module has recently been added. The MHD simulation is that made available for public use on the Johns Hopkins Turbulence database\footnote{available from:\\ 
\href{http://turbulence.pha.jhu.edu/Forced_MHD_turbulence.aspx}
{http://turbulence.pha.jhu.edu/Forced\_MHD\_turbulence.aspx.}}~\cite{2008JTurb...9...31L,2013Natur.497..466E}. 
The purpose of using different physical parameters and simulation frameworks is to test the capability of our model to describe the acceleration process under various turbulent regimes.

\subsection{2D decaying turbulence PIC simulation}

We initialize a 2D decaying turbulence PIC simulation with the following characteristics: domain size $N_x\times N_y = 10\,000^2$ cells, corresponding to physical size $L_x\times L_y = 1\,000^2 \, c^2/\omega_{\rm p}^2$, integrated over a time $T=5\,000\,\omega_{\rm p}^{-1}$. Here, $\omega_{\rm p} = \left(4\pi n_{\pm}e^2/m\right)^{1/2}$, with $n_{\pm}$ the initial (uniform) proper density of positrons/electrons and $e$ the elementary charge, represents the non-relativistic plasma frequency of one species, so $1/\sqrt{2}$ of the total plasma frequency. The cell size is $\delta x=\delta y=0.1\,c/\omega_{\rm p}$, and the time step $\delta t=0.099\,\omega_{\rm p}^{-1}$. The plasma is initialized with 10 particles per species per cell, sampled from a Maxwell-J\"uttner distribution function with a temperature $T_0=1\,\rm MeV$. The particle count ensures satisfactory statistics when making use of additional filtering as discussed below. Note also that, due to their large gyroradius, the high-energy particles studied here are less sensitive than thermal particles to small-scale field fluctuations. Periodic boundary conditions are used for both fields and particles in all directions.

Turbulence is excited as in Refs.~\cite{18Comisso,2019ApJ...886..122C,20Comisso}, {\it i.e.} a decaying turbulence initialized as a sum of plane waves. Here, we use 24 wavenumbers, with mean wavenumber $\langle k\rangle = (2\pi/L_x)\times 2.9$, corresponding to a stirring scale $\ell_{\rm c}\simeq L_x/2.9 \simeq 350\,c/\omega_{\rm p}$. The square root of the average of the squared wavenumbers give a similar estimate, $\langle k^2\rangle^{1/2} = (2\pi/L_x)\times 3.0$. Here, we will not distinguish the stirring scale from the coherence (or integral) scale; various definitions exist for the latter, which give values within a factor of the order of unity of $\ell_{\rm c}$. The modes initialized at time $0$ excite $\delta B_x$ and $\delta B_y$ fluctuations which are left to evolve freely with the plasma at time $t>0$. The mean field lies in the out-of-plane direction (along $z$). 

We define the magnetization parameter as
\begin{equation}
    \sigma\,=\,\frac{\langle B^2 \rangle}{4\pi w}\,,
    \label{eq:def-sigma}
\end{equation}
where $\langle B^2 \rangle$ is the mean-squared (coherent or turbulent) magnetic field and $w$ the plasma enthalpy density. For reference, $w \simeq 8 (n_+ + n_-) mc^2$ for an electron-positron pair plasma of $1\,\rm MeV$ temperature.
The magnetization associated with the mean-field component is $\sigma_0=1.6$. Since $\delta B/B_0 \simeq 2.8$, the turbulent magnetization is $\sigma_{\delta B} \simeq 13$.

\begin{figure}
\includegraphics[width=0.46\textwidth]{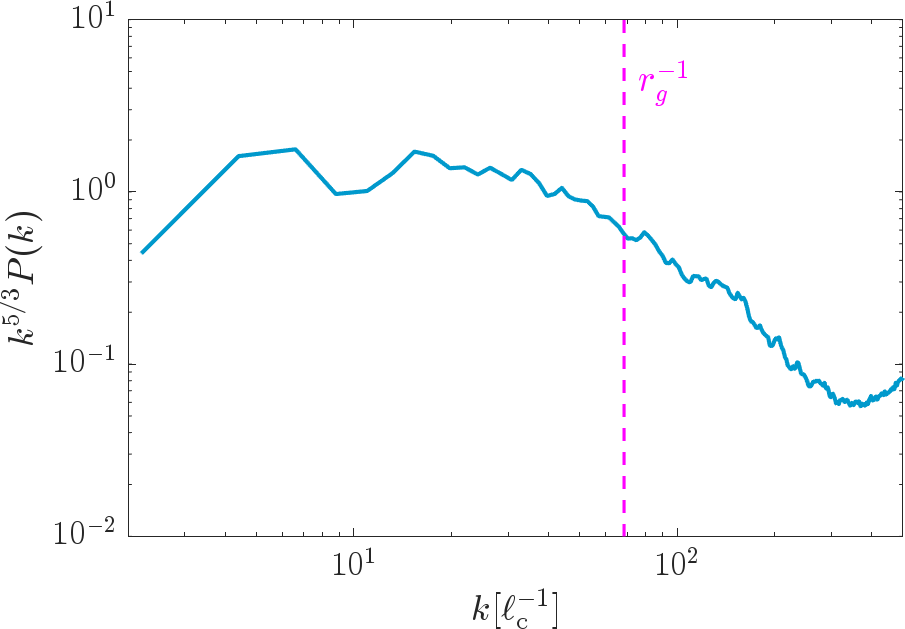}
 \caption{Power spectrum of magnetic fluctuations for the 2D decaying turbulence PIC simulation at $t \simeq 1800\,\omega_{\rm p}^{-1} \sim 5\ell_c/c$. The fuchsia vertical dashed line marks the scale $r_{\rm g}^{-1}$ for particles with Lorentz factor $\gamma=50$. Wavenumbers are given in units of the inverse stirring scale $\ell_{\rm c}^{-1}$; units on the $y-$axis are arbitrary.
 }
 \label{fig:pw-2ddec}
\end{figure}

In Fig.~\ref{fig:pw-2ddec}, we show the power spectrum of magnetic fluctuations as measured in this 2D decaying turbulence PIC simulation. It reveals a general scaling close to $k^{-5/3}$ at large scales, followed by a steeper spectrum characteristic of the dissipation range. This shape generally matches that observed in previous PIC simulations of decaying turbulence~\cite{18Comisso,2019ApJ...886..122C,20Comisso}. 

In that figure, we indicate by a dashed line the scale corresponding to the inverse gyroradius of particles with initial -- meaning, at the time $t=1500\,\omega_{\rm p}^{-1}\,\sim\,4\ell_{\rm c}/c$ at which we initiate the test -- Lorentz factor $\gamma\sim 50$, which we follow in order to compare the model to the data using the method described earlier\footnote{We select here a range in $\gamma$, not $\gamma'$, but this does not significantly influence our results, as we have explicitly checked for the 2D PIC driven turbulence and the MHD simulations discussed further below.}. We recall that this model assumes $r_{\rm g}\ll\ell_{\rm c}$, hence $r_{\rm g}$ cannot be made arbitrarily larger. However, it cannot be made arbitrarily small either, otherwise the particle gyroradius will lie out of the range of the inertial (non-dissipative) spectrum. On small spatial scales, corresponding to gyroradii of particles with energies in the thermal part of the spectrum -- $\gamma\sim 10$ -- particle energization is furthermore mostly controlled by parallel electric fields, as recalled above~\cite{18Comisso,2019ApJ...886..122C}. We thus conclude that Lorentz factors in the range $\sim 20-60$ provide a reasonable compromise to test the theoretical model of non-resonant acceleration. In the present case, the effective rigidity $2\pi r_{\rm g}/\ell_{\rm c}$ of particles with Lorentz factor $\gamma\sim50$ is of the order of $0.1$, in the range anticipated earlier.

\begin{figure}
\includegraphics[width=0.46\textwidth]{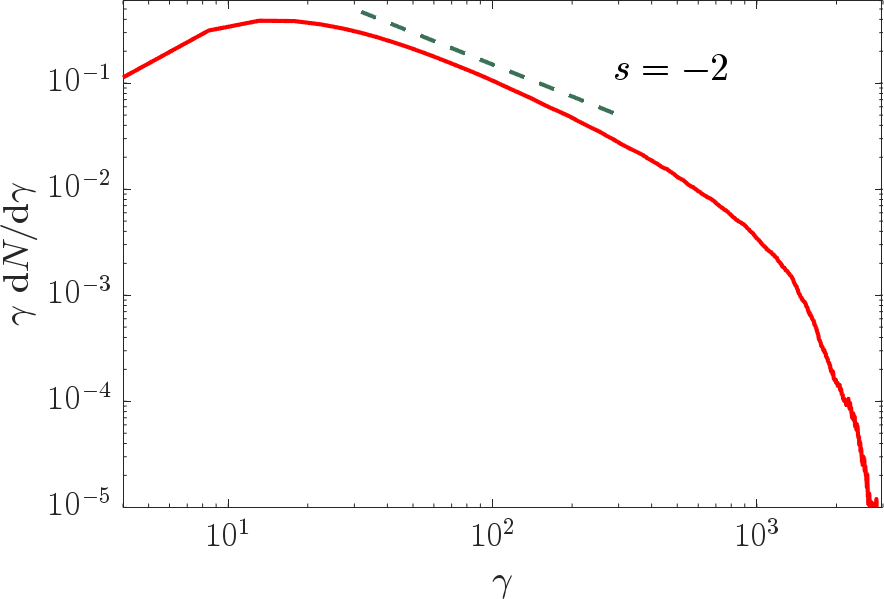}
 \caption{Energy distribution of the particles in the 2D decaying turbulence PIC simulation at $t \simeq 1800\, \omega_{\rm p}^{-1} \sim 5\ell_c/c$. The powerlaw tail emerges at $\gamma\gtrsim 20$ up to $\gamma\sim 10^3$, with spectral index $s\simeq -2$, defining $s$ through ${\rm d} N/{\rm d}\gamma \propto \gamma^s$. 
 }
\label{fig:sp-2ddec} 
\end{figure}

In Fig.~\ref{fig:sp-2ddec}, we show the energy spectrum of particles in this 2D decaying turbulence simulation, at time $t\simeq 5\ell_{\rm c}/c$. The powerlaw tail, with index $s\simeq -2$, extends from Lorentz factors $\gamma\sim 20$ up to $\gamma\sim 10^3$, at which point the gyroradius of particles becomes comparable to the maximal scale of the turbulent cascade, implying less efficient acceleration at larger energies. 

We now turn to the comparison between the model predictions and the PIC simulation. The trajectories of a sample/subset of particles are recorded from $t=1500\,\omega_{\rm p}^{-1}\,\sim\,4\ell_{\rm c}/c$ up to $5000\,\omega_{\rm p}^{-1}\sim14\ell_{\rm c}/c$. 
The initial time ensures that the turbulence has had time to cascade down to small scales by the time the test starts. 
In the PIC simulations, both time and space derivatives are calculated using simple first-order differences, {\it i.e.} from one cell to the next (or one step to the next for time). The time derivatives are smoothed through 16 repeated applications of binomial filtering. The spatial derivatives are computed from fields that also underwent 16 successive applications of binomial filtering. This helps eliminate shot noise on the scale of the mesh (here, $\sim 0.1\,c/\omega_{\rm p})$ that would otherwise pollute the reconstruction of derivatives which, as discussed before, are meant to be calculated on scales significantly larger than the grid size.

\begin{figure}
\includegraphics[width=0.48\textwidth]{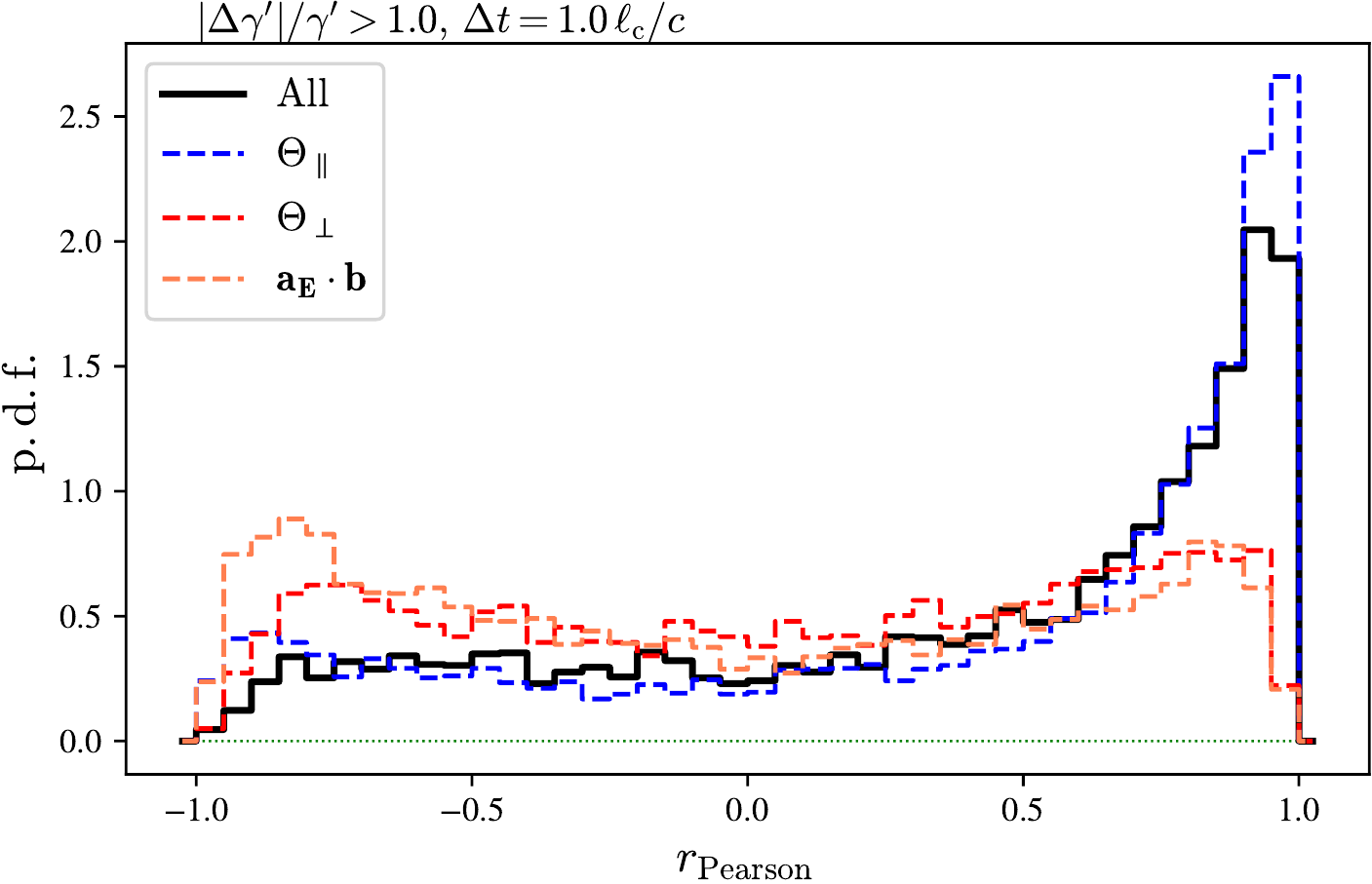}
 \caption{Histogram of correlation coefficients between the expected and observed evolution of $\gamma'$ along chunks of test particle trajectories with initial $25<\gamma<50$, for the 2D decaying turbulence PIC simulation. The chunks are selected at random among the whole of test particle histories, provided they fulfill the following criteria: the duration $\Delta t \simeq 1\,\ell_{\rm c}/c$ and the energy change within that time interval verifies $\left\vert\Delta\gamma'/\gamma'\right\vert>1$. 
 This histogram shows that the parallel shear $\Theta_\parallel$ contribution, and more generally the non-resonant model as described by Eq.~(\ref{eq:Rco-evol}), match relatively well the observed variations.
 \label{fig:hist-ck-2ddec} }
\end{figure}

\begin{figure}
\includegraphics[width=0.48\textwidth]{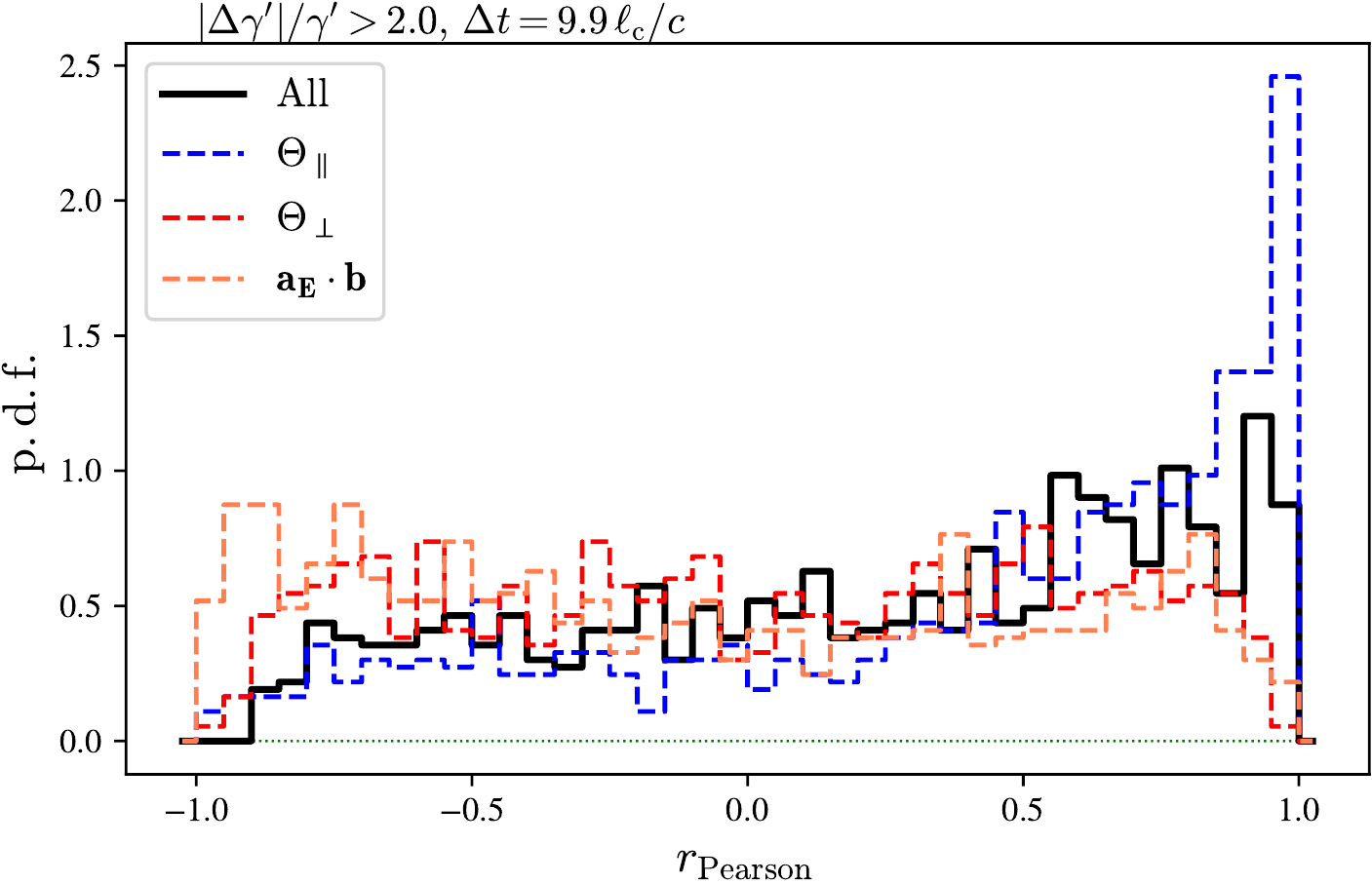}
 \caption{Same as Fig.~\ref{fig:hist-ck-2ddec} (2D decaying turbulence PIC simulation), now considering the whole trajectory of each test particle with $\left\vert\Delta\gamma'/\gamma'\right\vert>2$. 
 \label{fig:hist-all-2ddec} }
\end{figure}

The test has been carried out for two typical durations: $\Delta t \simeq 1\,\ell_{\rm c}/c$, see Fig.~\ref{fig:hist-ck-2ddec} and $\Delta t \simeq 10 \,\ell_{\rm c}/c$, 
see Fig.~\ref{fig:hist-all-2ddec}. Those figures present the probability density function (p.d.f.) of the correlation coefficients $r_i$, as defined in Eq.~(\ref{eq:Pearson-c}). To construct the histogram shown in Fig.~\ref{fig:hist-ck-2ddec}, we have selected at random, for each test particle, chunks of trajectories in which the energy of the particle changes by an amount at least equal to unity, {\it i.e.} $\Delta\gamma'/\gamma'\geq 1$ with $\Delta\gamma'={\rm max}\left(\gamma'\right)-{\rm min}\left(\gamma'\right)$. We typically use $10^4$ test particles to construct such a histogram; each test particle history, extending over $\simeq 10\ell_{\rm c}$, is sampled at most $10$ times to obtain a chunk of extent $1\,\ell_{\rm c}/c$. In Fig.~\ref{fig:hist-all-2ddec}, we integrate over the whole trajectory of each test-particle, provided $\vert\Delta\gamma'/\gamma'\vert\geq 2$. 

Figure~\ref{fig:hist-ck-2ddec} indicates a genuinely positive degree of correlation for the contribution of the $\Theta_\parallel$ force terms, and similarly when all contributions are summed together as in Eq.~(\ref{eq:Rco-evol}). More specifically, to plot the probability density of the correlation coefficients for one force contribution, we use Eq.~(\ref{eq:Rco-evol}) but set the contributions of the other two terms to zero. This figure suggests that neither the force term $\Theta_\perp$ nor $\boldsymbol{a_E}\cdot\boldsymbol{b}$ appear to 
contribute strongly to the evolution of the particle energy. The dominance of $\Theta_\parallel$ is a common trait to our PIC simulations, which will also hold in 3D as discussed further on. 

The trend observed in Fig.~\ref{fig:hist-all-2ddec} is similar. The level of noise is larger in that figure, because we can select only one trajectory for each test-particle instead of a number of distinct time intervals, and because our stronger constraint on the amount of energy variation within the interval limits further the number of test particles that are selected for the test.

We note that the above figures and results are relatively insensitive to the choice of the threshold of energy variation $\vert\Delta\gamma'/\gamma'\vert$, as we have verified. It is also somewhat insensitive to the duration of the interval that we consider. The latter must be large enough, obviously, to accommodate a large number of gyroperiods, since the model considers only contributions from scales larger than $r_{\rm g}$.

\subsection{2D forced turbulence PIC simulation}

We now analyze a 2D driven turbulence PIC simulation with characteristics similar to that for the decaying turbulence scenario: domain size $N_x\times N_y = 10\,000^2$ cells, corresponding to physical size $L_x\times L_y = 1\,000^2\, c^2/\omega_{\rm p}^2$, integrated over time $T=5\,000\,\omega_{\rm p}^{-1}$; the cell and step size are, as before, $\delta x = \delta y = 0.1\,c/\omega_{\rm p}$ and $\delta t = 0.099\,\omega_{\rm p}^{-1}$. 
The initial magnetizations are the same as for the decaying turbulence scenario, $\sigma_0\simeq 1.6$ and  $\sigma_{\delta B}\simeq 13$. 

Turbulence is excited using a Langevin antenna scheme~\cite{2014CoPhC.185..578T}, in a way similar to the implementation of Refs.~\cite{17Zhdankin,2018ApJ...867L..18Z,2019PhRvL.122e5101Z,2020ApJ...893L...7W}. For the present 2D simulations, we excite external current fluctuations along the mean magnetic field ($z-$axis) only, with wavemodes oriented in the $(x,y)$ plane. We use 24 modes, with mean wavenumber $\langle k\rangle = (2\pi/L_x)\times 2.9$ (and similar $\langle k^2\rangle^{1/2}$), implying $\ell_{\rm c}\sim 350\,c/\omega_{\rm p}$ as before. Those external currents generate $\delta B_x$ and $\delta B_y$. We then tune the amplitude of the antenna to reproduce the chosen initial turbulent magnetization. The Langevin antenna is also characterized by a real frequency $\omega_0$ and a damping term $\Gamma_0$. We found it useful to set the real frequency to low values, in practice $\omega_0\approx0$, in order to avoid excessive heating of the plasma at early times, caused by the rapid generation of non-MHD electric fields on large scales. Regarding the damping term, we tune it in order to ensure that the auto-correlation time of the turbulent magnetic field matches roughly $\ell_{\rm c}/c$; in practice, we set $\Gamma_0\simeq 0.6 \langle k\rangle\,c$. 

\begin{figure}
\includegraphics[width=0.46\textwidth]{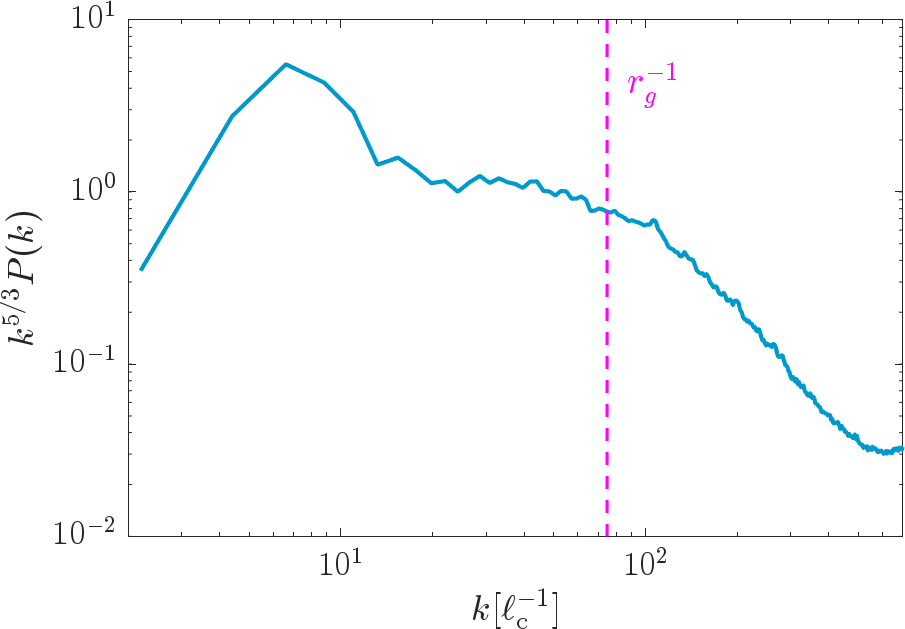}
 \caption{Power spectrum of magnetic fluctuations for the 2D forced turbulence PIC simulation at $t \simeq 1800\, \omega_{\rm p}^{-1} \sim 5\ell_c/c$. The fuchsia vertical dashed line marks the scale $r_{\rm g}^{-1}$ for particles with Lorentz factor $\gamma=50$.
 }
 \label{fig:pw-2dfor}
\end{figure}

The power spectrum of magnetic fluctuations shown in Fig.~\ref{fig:pw-2dfor} reveals a shape similar to that seen in the decaying turbulence case, with a (roughly) $k^{-5/3}$ generic scaling over the inertial domain, followed by the steeper dissipative range at kinetic scales. The spectrum amplitude is more pronounced at the stirring scale, a trend which is characteristic of forced turbulence PIC simulations, if one compares Refs.~\cite{2019ApJ...886..122C} and ~\cite{2018MNRAS.474.2514Z}. This is expected insofar turbulence is continuously injected at the stirring scale in forced turbulence, while the spectrum moves in time to larger $k$ in decaying turbulence.  

As in the decaying turbulence scenario, we indicate with a dashed line the inverse gyroradius of particles with Lorentz factor $\gamma\sim 50$. Again, their effective rigidity is of the order of $0.1$, which falls in the right range to test the non-resonant acceleration model. 

\begin{figure}
\includegraphics[width=0.46\textwidth]{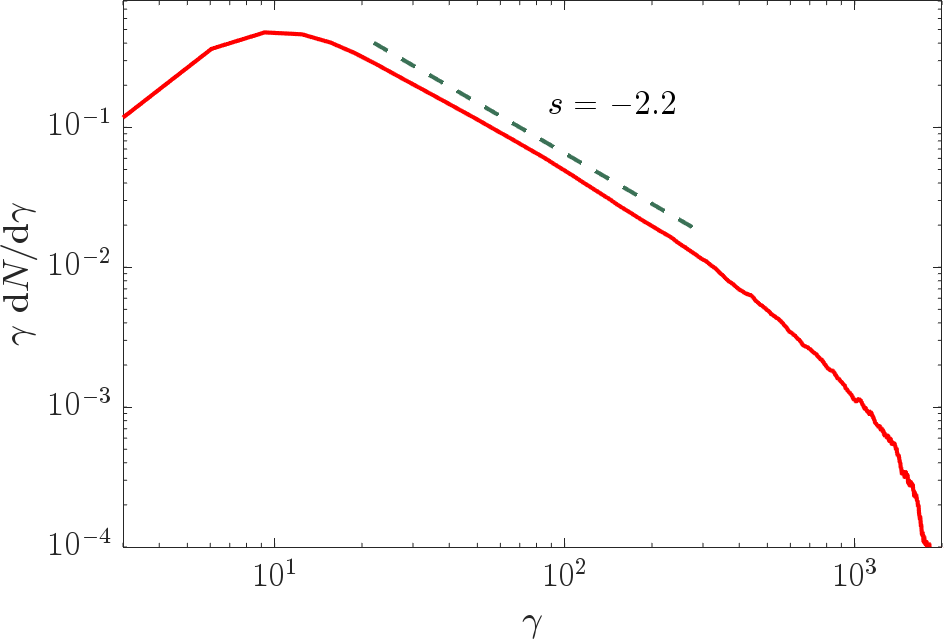}
 \caption{Energy distribution of the particles in the 2D forced turbulence PIC simulation at $t \simeq 1800\,\omega_{\rm p}^{-1} \sim 5\ell_c/c$. A powerlaw tail is clearly seen, extending from $\gamma\sim 10$ up to $\gamma\sim 10^3$, with spectral index $s\simeq -2$, defining $s$ through ${\rm d} N/{\rm d}\gamma \propto \gamma^s$. The test particles that we study, with $25<\gamma<50$, are located in the powerlaw tail.
 } 
 \label{fig:sp-2dfor}
\end{figure}

In Fig.~\ref{fig:sp-2dfor}, we plot the particle energy distribution, which reveals a powerlaw tail extending from $\gamma\sim10$ up to $\gamma\sim10^3$, as for the decaying turbulence scenario. The best-fitting spectral index, $s\simeq -2.2$, is also close to that found previously. We note that in forced turbulence simulations, the spectrum evolves slowly in time, as the energy that is continuously injected in the simulation maintains $\delta B/B$ (and, to a lesser degree, the overall magnetization) at values not far from its initial state, thereby guaranteeing that acceleration can proceed at all times. In decaying turbulence simulations, the drop in magnetization associated with magnetic dissipation within $\sim 5-10\,\ell_{\rm c}/c$ implies that acceleration becomes much slower, so that the spectrum essentially freezes on those timescales~\cite{18Comisso,2019ApJ...886..122C,20Comisso}.

\begin{figure}
\includegraphics[width=0.48\textwidth]{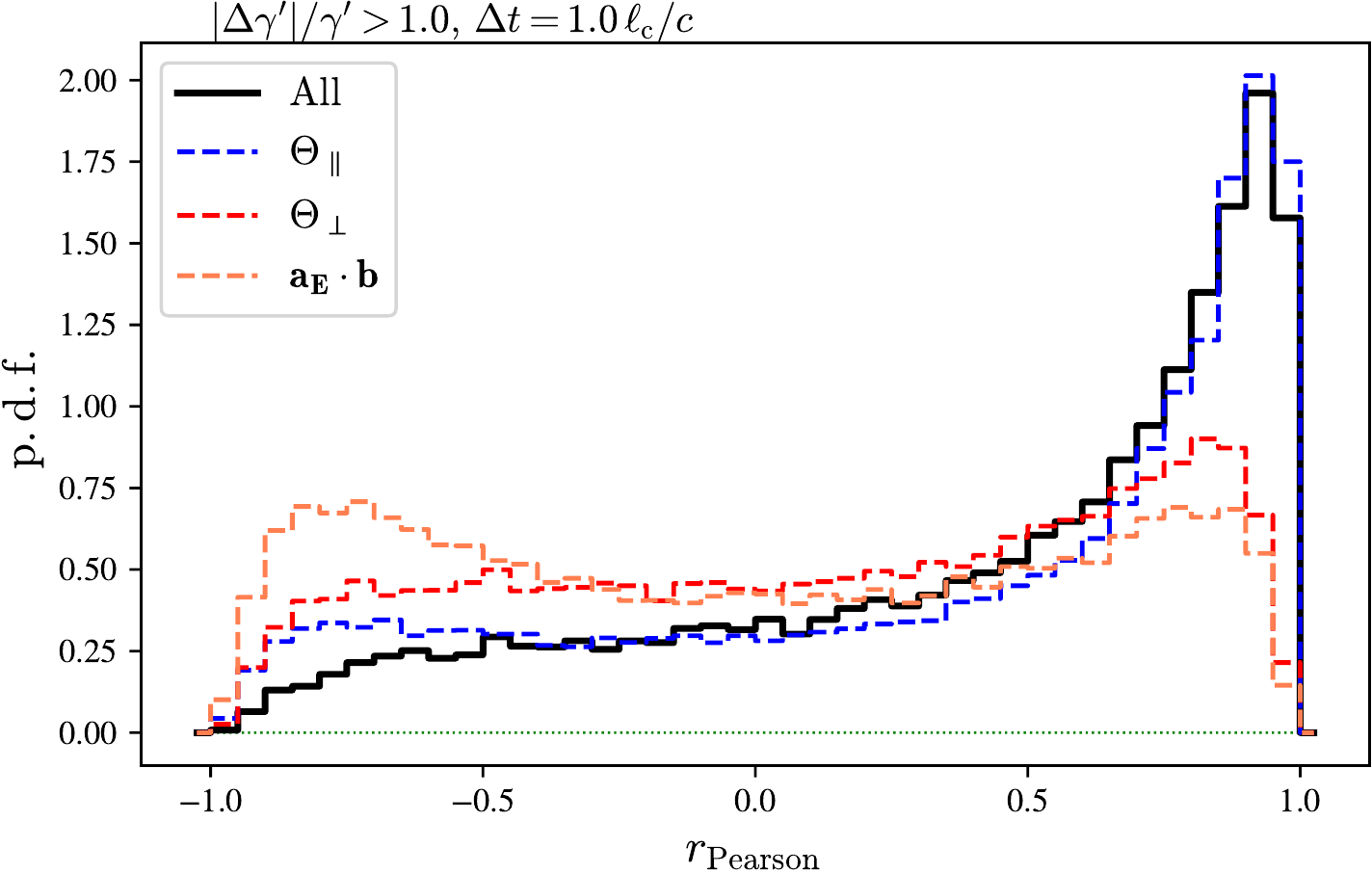}
 \caption{Histogram of correlation coefficients between the expected and observed evolution of $\gamma'$ along chunks of test particle trajectories with initial $25<\gamma<50$, for the 2D forced turbulence PIC simulation. This histogram shows that the non-resonant model provides a satisfactory match to the observed variations, and that the parallel shear term $\Theta_\parallel$ provides the dominant contribution to the force terms.
 \label{fig:hist-ck-2dfor} }
\end{figure}

\begin{figure}
\includegraphics[width=0.48\textwidth]{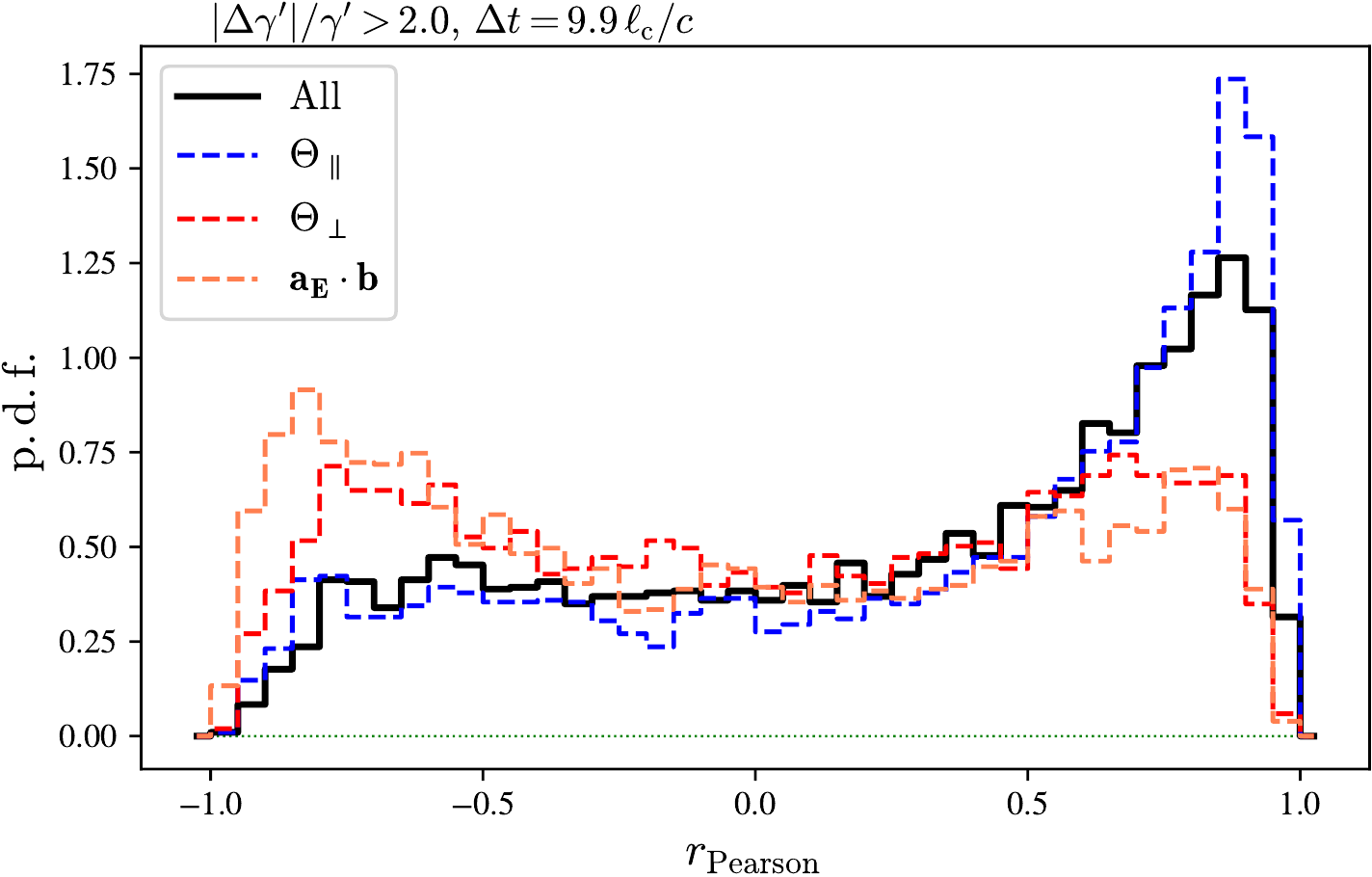}
 \caption{Same as Fig.~\ref{fig:hist-ck-2dfor} (2D forced turbulence PIC simulation), now considering the whole trajectory of each test particle.
 \label{fig:hist-all-2dfor} }
\end{figure}

To reconstruct the probability density distributions of the correlation coefficients between observed and reconstructed histories, we follow test particles from $t=1\,500\,\omega_{\rm p}^{-1}\sim4\ell_{\rm c}/c$ up to $5\,000\,\omega_{\rm p}^{-1}\sim 14\ell_{\rm c}/c$, as for decaying turbulence. We present those probability densities in Figs.~\ref{fig:hist-ck-2dfor} and \ref{fig:hist-all-2dfor}. The parameters (duration, amount of variation of the energy) are the same as in the decaying turbulence scenario. We observe a similar trend, namely the non-resonant model captures fairly well the observed energy histories, and $\Theta_\parallel$ provides the dominant contribution among the three force terms. The perpendicular contribution $\Theta_\perp$ and the inertial term do not show such significant degrees of correlation, although that of $\Theta_\perp$ is skewed towards positive values, at least for short $\Delta t \simeq 1\,\ell_{\rm c}/c$ timescales.

Generally speaking, the degree of agreement between model and simulations appears more satisfactory for the present forced turbulence scenario than for the decaying one. In that respect, we note that the shape of the spectrum can impact this comparison in the following way. The power spectrum of the forced simulation shows a larger amplitude on the smallest $k-$modes, meaning on the largest length scales, than the decaying turbulence one, all things being considered equal. This difference can be read off Figs.~\ref{fig:pw-2ddec} and \ref{fig:pw-2dfor}, but it is actually more pronounced at later times, since the power spectrum of the decaying turbulence shifts to larger $k$ as time progresses.  This implies that, on the whole, particles experience a turbulence on larger scales in the forced turbulence case than in the decaying turbulence one, as measured relatively to their gyroradius. Since the model works to order $r_{\rm g}/\ell_{\rm c}$, this larger degree of agreement is therefore not unexpected, at least at a qualitative level.

\subsection{3D forced turbulence PIC simulation}

\begin{figure}[h!]
\includegraphics[width=0.48\textwidth]{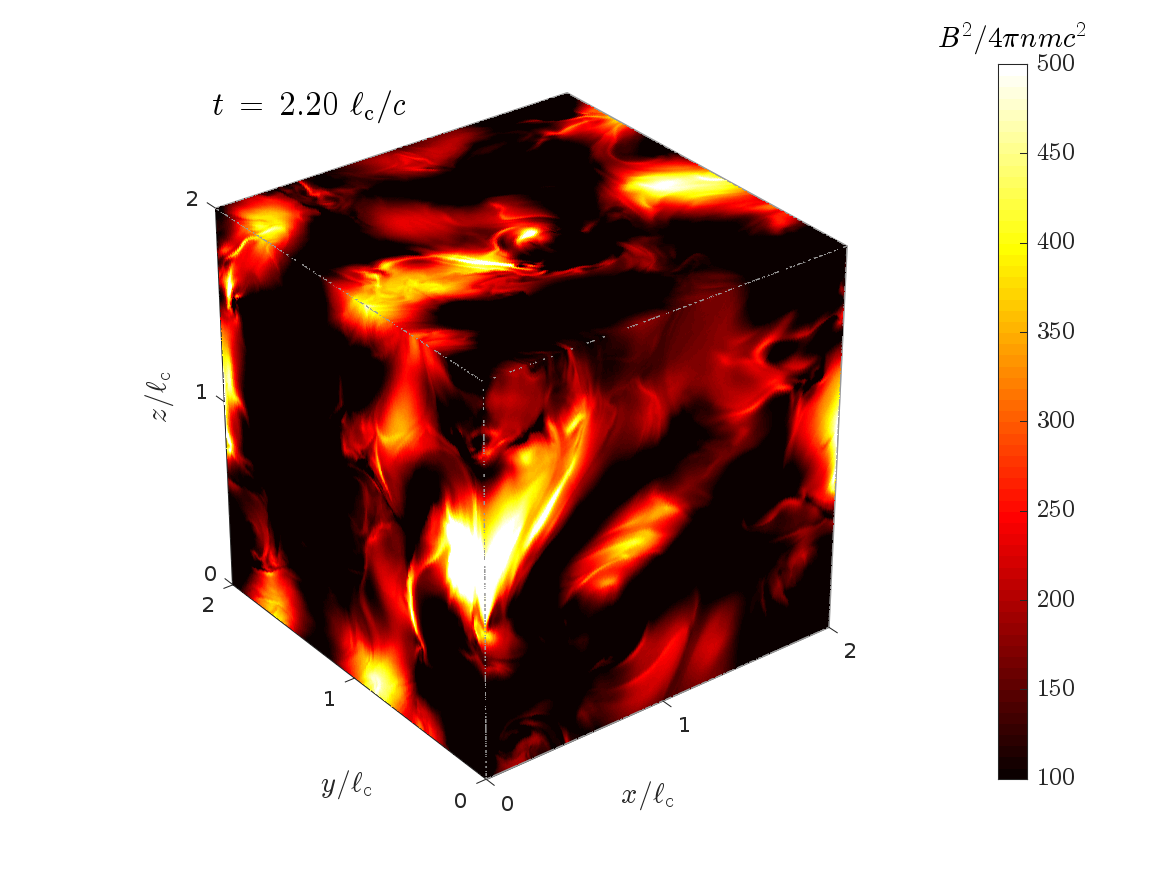}
\includegraphics[width=0.48\textwidth]{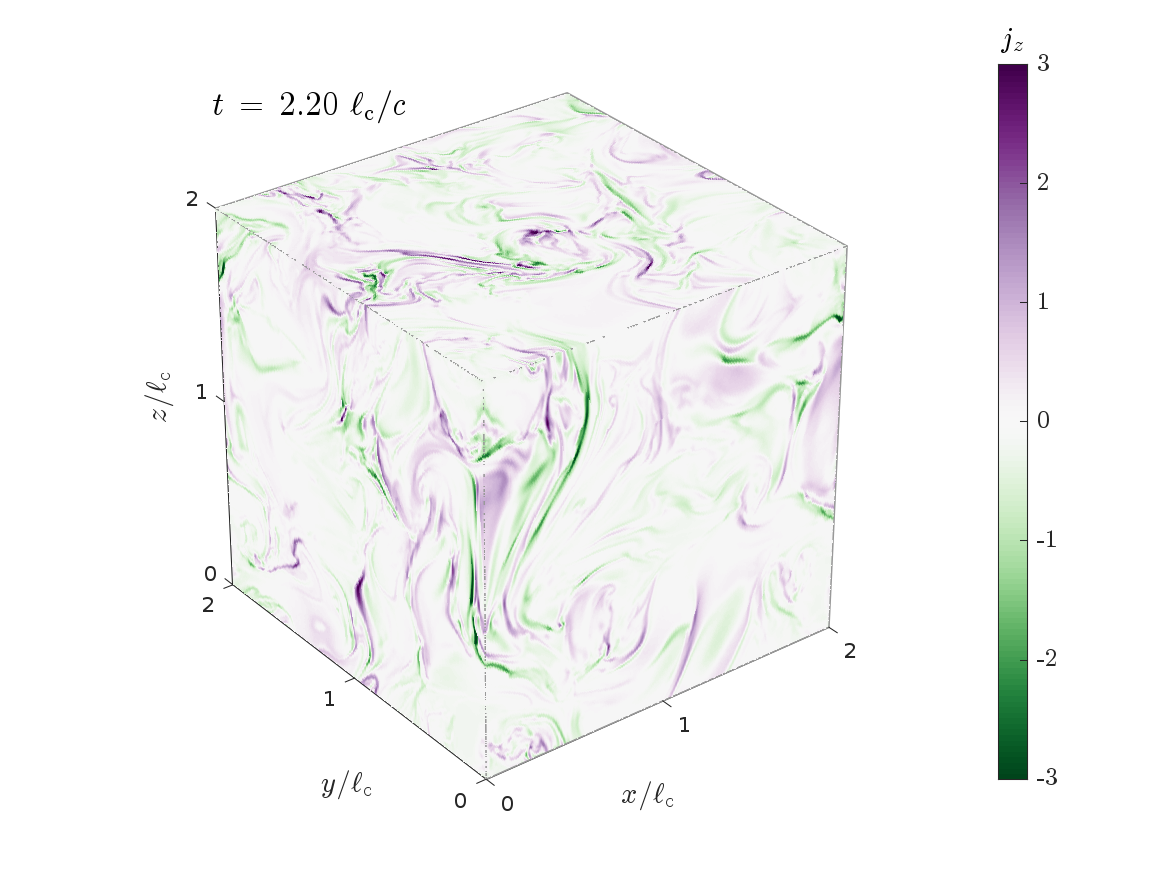}
\includegraphics[width=0.48\textwidth]{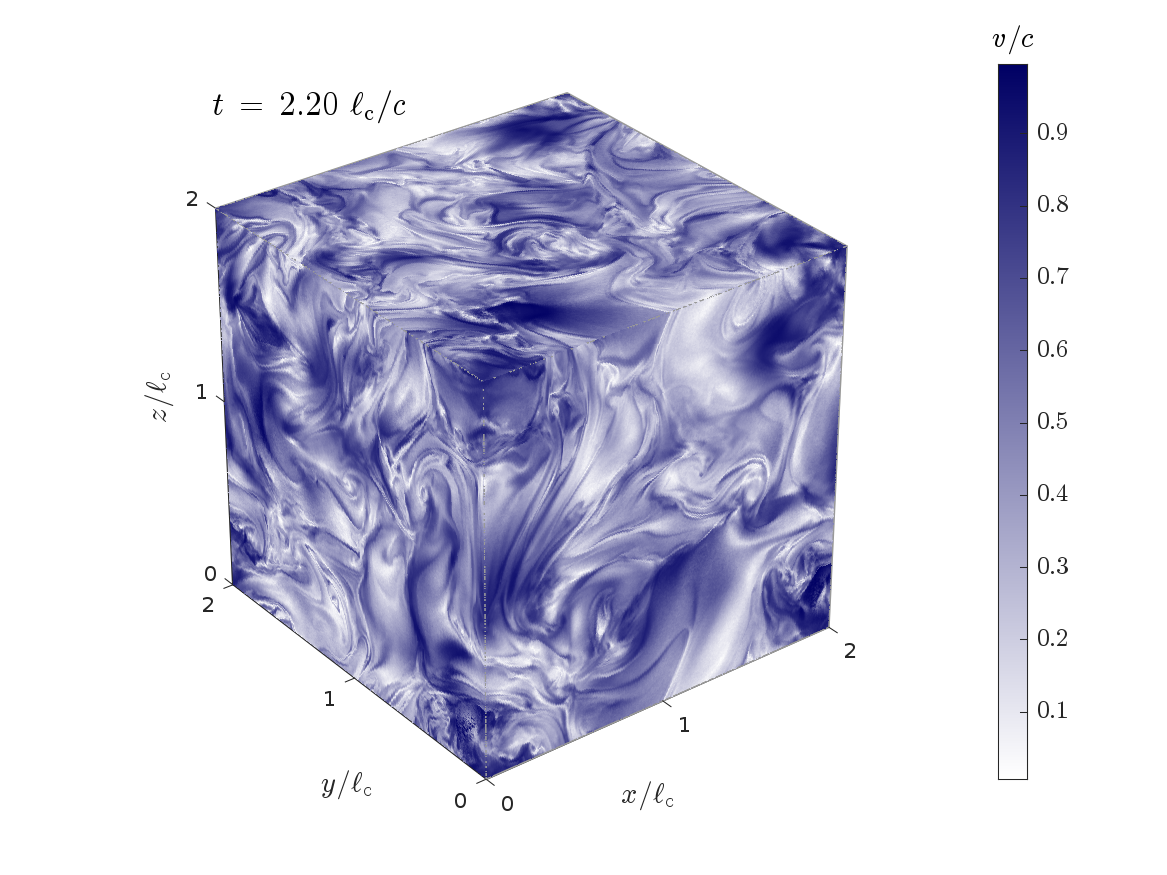}
 \caption{Volume rendering of the 3D PIC simulation, showing the magnetic energy density (in units of plasma rest-mass energy, top panel), the current density component along the mean magnetic field component (middle panel) and the mean plasma velocity (in units of $c$, bottom panel).
 \label{fig:3Dplot} }
\end{figure}

We now turn to 3D simulations. We have first performed a 3D forced turbulence PIC simulation with domain size $N_x\times N_y\times N_z = 1\,080^3$ cells, corresponding to physical size $L_x\times L_y\times L_z = 540^3\, c^3/\omega_{\rm p}^3$, integrated over $5\,000$ time steps, corresponding to a time $t \simeq 1\,500\,\omega_{\rm p}^{-1}$; the mesh size is now $\delta x=\delta y=\delta z= 0.5\,c/\omega_{\rm p}$, $\delta t =0.495\,\omega_{\rm p}^{-1}$ and we use 15 particles per species per cell. This choice of parameters is motivated by the need to optimize the execution time, while avoiding excessive shot noise associated with the number of macro-particles per skin depth volume. When measured in terms of the total relativistic plasma frequency, $\Omega_{\rm p}\,=\,\left[ 4\pi(n_++n_-)e^2/w\right]^{1/2}$, the mesh size reads $\delta x \simeq 0.25 \,c/\Omega_{\rm p}$; given that the plasma further heats with time in the turbulence, this provides a relatively fair sampling of the skin depth volume. Figure~\ref{fig:3Dplot} offers a general view on the simulation at time $t\simeq 600\,\omega_{\rm p}^{-1}$: magnetic energy density (top panel), current density component along the mean field direction (middle panel) and plasma bulk velocity (bottom panel).

The initial mean field magnetization is $\sigma_0=1.6$ as before, while $\sigma_{\delta B} \simeq 8$. The forced turbulence is excited using the same Langevin antenna scheme as in 2D, with the following parameters: in 3D, we generate 24 modes of external current density fluctuations along $x$, along $y$ and along $z$ separately, with mean wavenumbers $\langle k\rangle = 2\pi/L_{\rm max}\times 2.$; for reference, $\langle k^2\rangle^{1/2} \simeq 2\pi/L_{\rm max}\times 2.$ as well, guaranteeing a stirring scale $\ell_{\rm c}\simeq L_{\rm max}/2 \simeq 270\,c/\omega_{\rm p}$. We use a real frequency $\omega_0=0$ to avoid the generation of external electric fields in the forcing scheme, and a damping term $\Gamma_0=0.4 \langle k\rangle c$.

\begin{figure}
\includegraphics[width=0.46\textwidth]{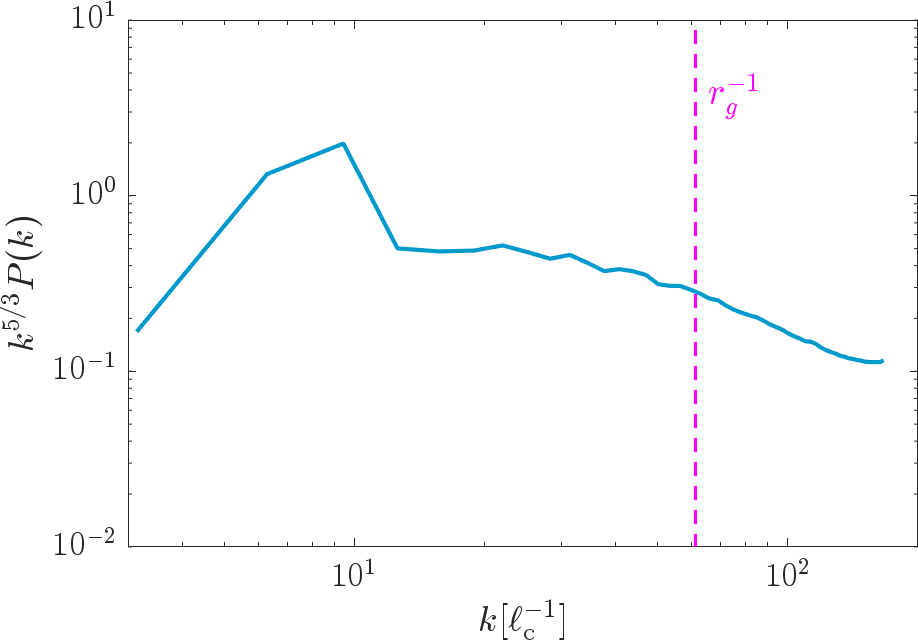}
 \caption{Power spectrum of magnetic fluctuations for the 3D forced turbulence PIC simulation at $t \sim 600 \omega_{\rm p}^{-1} \sim \ell_c/c$. The fuchsia vertical dashed line marks the scale $r_{\rm g}^{-1}$ for particles with Lorentz factor $\gamma=50$. The fall-off of the spectrum in the dissipative range is not as prominent as in 2D due to the data rebinning used (see main text).
 }
 \label{fig:pw-3dfor}
\end{figure}

\begin{figure}
\includegraphics[width=0.46\textwidth]{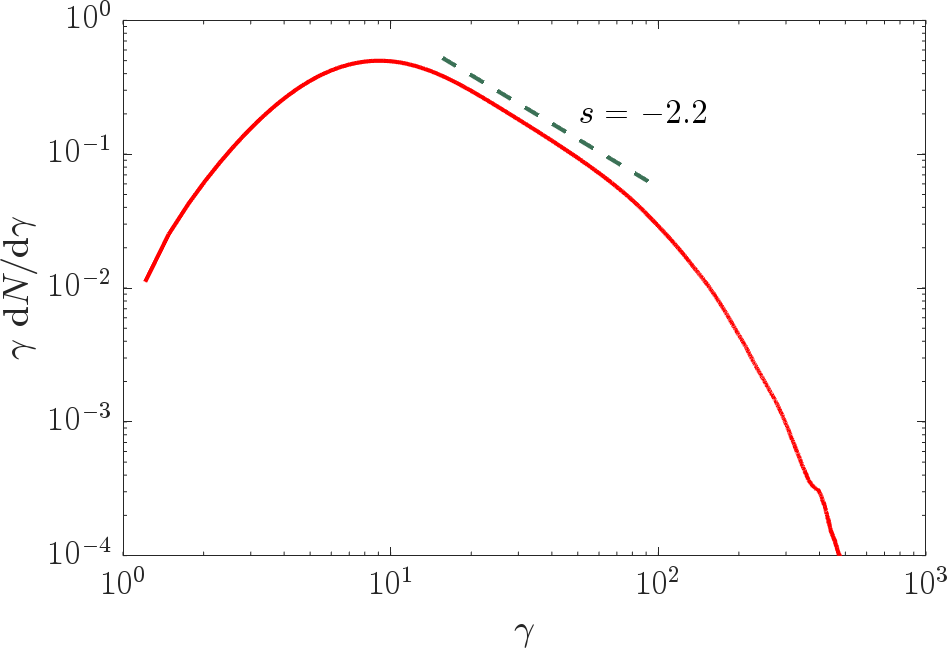}
 \caption{Energy distribution of the particles in the 3D forced turbulence PIC simulation at $t \sim 600 \omega_{\rm p}^{-1} \sim 2\ell_c/c$. A powerlaw tail with spectral index $s\simeq -2.2$ is clearly seen, extending from $\gamma\sim 10$ up to $\gamma\sim 100$.
 }
 \label{fig:sp-3dfor}
\end{figure}

As before, we present in Fig.~\ref{fig:pw-3dfor} the power spectrum of magnetic fluctuations and in Fig.~\ref{fig:sp-3dfor} the energy distribution of particles,  at $t\simeq 600\,\omega_{\rm p}^{-1}\sim 2.2\, \ell_{\rm c}/c$. To compute the 3D power spectrum (and preserve memory usage), the field values have been rebinned by ten, so that the minimum length scale plotted is $10\delta_x= 5\,c/\omega_{\rm p}$. Consequently, the power spectrum shown in Fig.~\ref{fig:pw-3dfor} lacks data at large wavenumbers (in the dissipative range); it covers about two decades, even though the grid size contains $1\,080$ cells along each its axis.

That timescale $t\sim 2.2 \ell_{\rm c}/c$ is shorter than that used in 2D PIC simulations for plotting purposes, because of the shorter duration of that 3D simulation. Consequently, the peak amplitude associated with the externally injected energy appears more prominent in the 3D simulation, and the powerlaw tail of the energy distribution has not yet reached the maximum energy fixed by the coherence length, of the order of several hundreds here.

\begin{figure}
\includegraphics[width=0.48\textwidth]{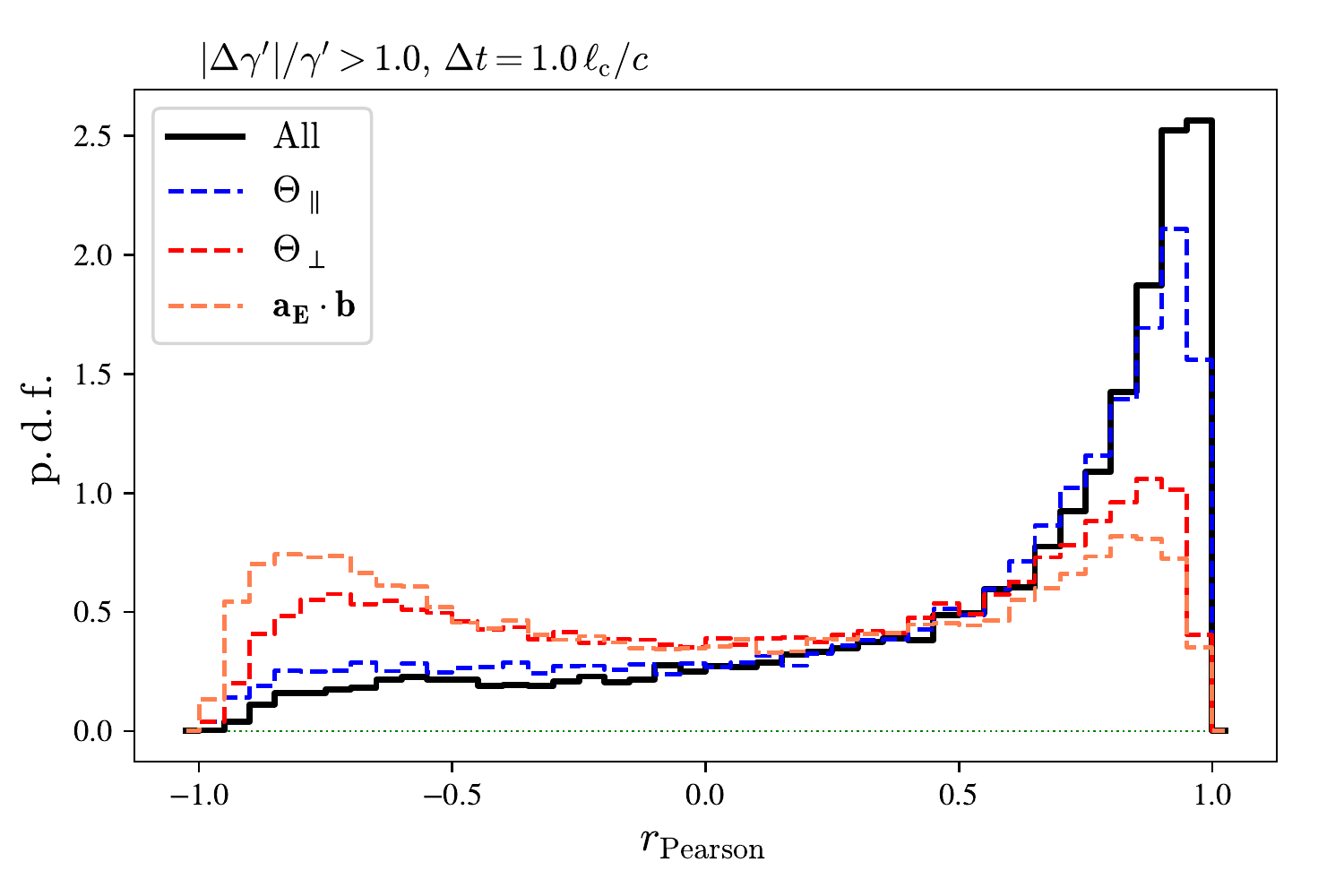}
 \caption{Histogram of correlation coefficients between the expected and observed evolution of $\gamma'$ along chunks of test particle trajectories with initial $25<\gamma<50$, for the 3D forced turbulence PIC simulation. This histogram shows that the combination of all force terms provides a good match to the observed variations. Among the three force terms, the parallel shear $\Theta_\parallel$ provides the dominant contribution.
 \label{fig:hist-ck-3dfor} }
\end{figure}

To compute the probability density histograms of the correlation coefficients $r_i$, we have followed the test particle trajectories from $t\simeq 500\,\omega_{\rm p}^{-1}\sim 2\,\ell_{\rm c}/c$ up to $1\,500\,\omega_{\rm p}^{-1}\sim 6\,\ell_{\rm c}/c$, which marks the duration of the simulation. As anticipated, the p.d.f. is sharply peaked around $+1$ for this 3D simulation, indicating a nice match between the energy variations predicted by the model and those observed in the simulation. The parallel compression term $\Theta_\parallel$ provides as before the leading contribution; the p.d.f. of the perpendicular force term is slightly biased toward positive values, as for the 2D forced turbulence simulation, while the inertial term does not show any clear signature, as in 2D. Interestingly, the correlation appears slightly enhanced when all force terms are taken together as in Eq.~(\ref{eq:Rco-evol}), than when they are taken one by one, at least for the case in which intervals of duration $1\ell_{\rm c}/c$ are examined.

\begin{figure}
\includegraphics[width=0.48\textwidth]{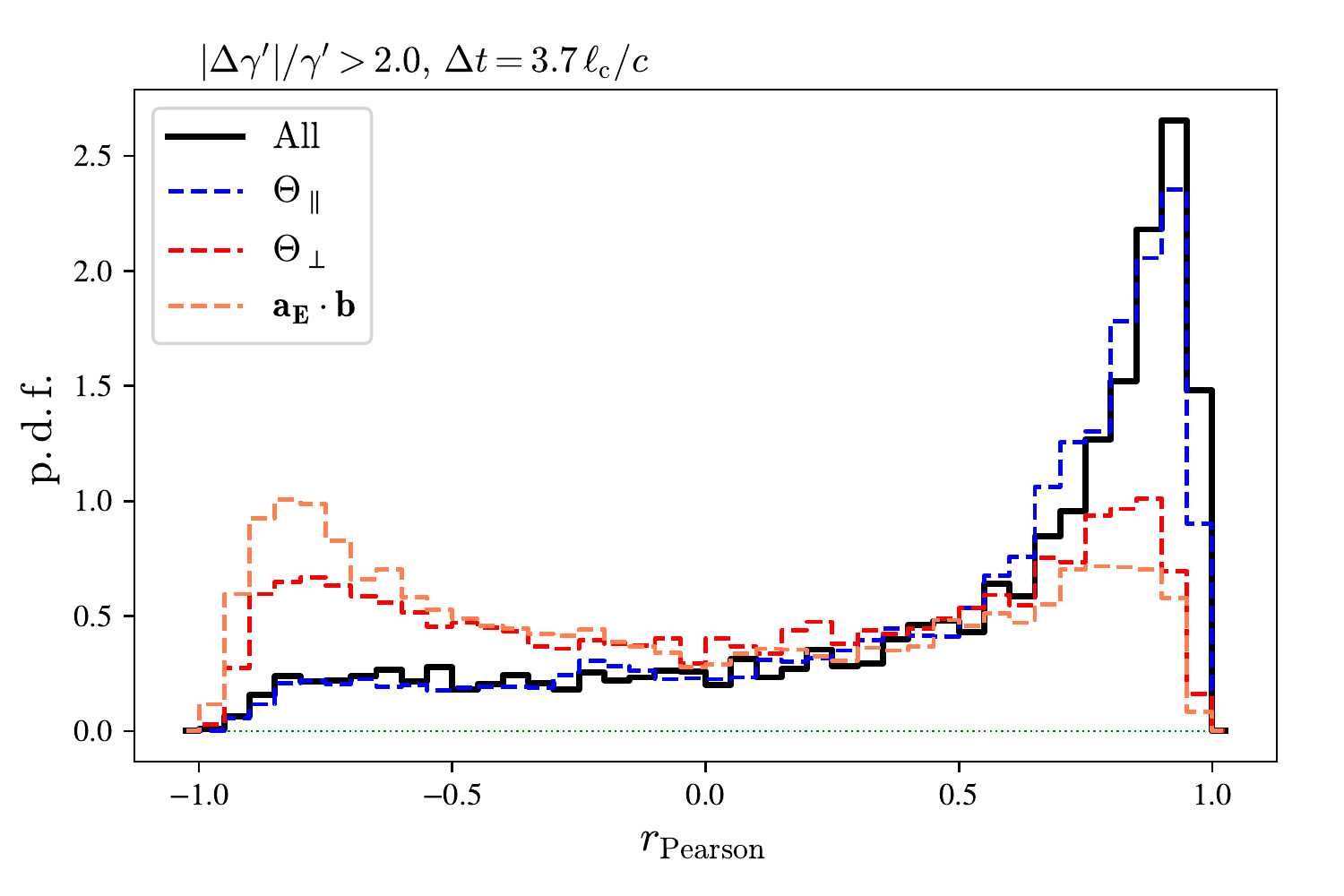}
 \caption{Same as Fig.~\ref{fig:hist-ck-3dfor} (3D forced turbulence PIC simulation), now integrating over the whole trajectory for each test particle.
 \label{fig:hist-all-3dfor} }
\end{figure}

\begin{figure}
\includegraphics[width=0.48\textwidth]{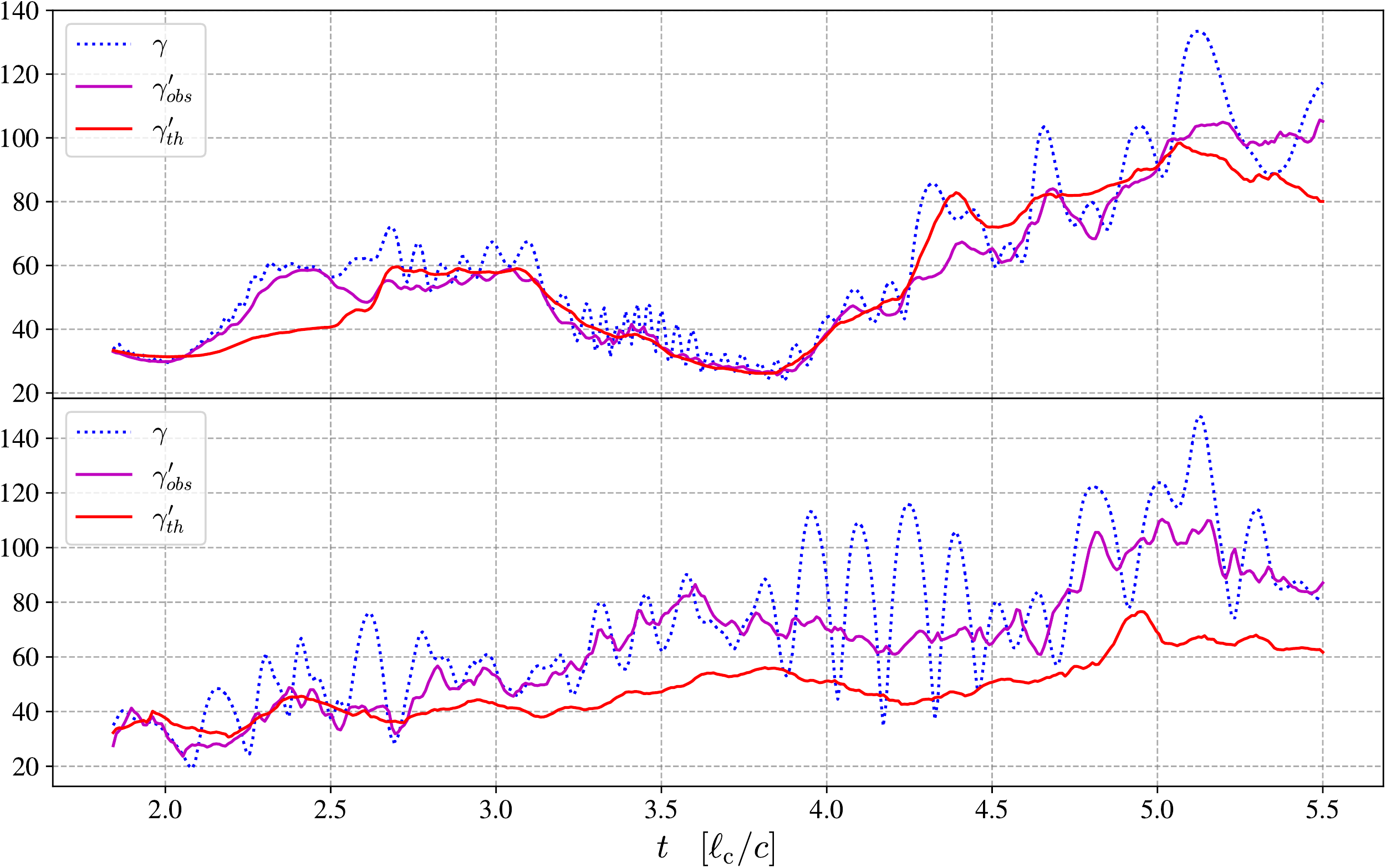}
 \caption{Example of the time evolution of the energy for two test particles, in the 3D forced turbulence simulation. In dashed blue: the energy of the particle as measured in the simulation frame; in solid purple: the particle Lorentz factor $\gamma'_{\rm obs}$ in the $\mathcal R_{\slashed E}$ frame; in solid red: the Lorentz factor $\gamma'_{\rm th}$, as reconstructed using Eq.~(\ref{eq:Rco-evol}). 
 }
 \label{fig:traj-all-3dfor}
\end{figure}

In Fig.~\ref{fig:traj-all-3dfor}, we present the energy evolution of two test particles, which are fair representatives of their parent population. The dotted blue line shows the evolution of $\gamma(t)$, {\it i.e.}, the Lorentz factor of the particle as measured in the simulation frame. It displays characteristic oscillations associated with the gyromotion of the particle around magnetic field lines that move at velocity $\boldsymbol{v_E}$: depending on the phase of that gyromotion, the particle motion is aligned or anti-aligned with $\boldsymbol{v_E}$, leading to a larger or smaller apparent energy in the simulation frame, see  also Refs.~\cite{2020ApJ...893L...7W,2020PhRvD.102b3003D}. The period of those oscillations thus provides an estimate of $2\pi r_{\rm g}/c$, which takes different values at different times, depending on the strength of the magnetic field and of the particle energy. The solid purple line shows the evolution of $\gamma'_{\rm obs}(t)$, in the frame $\mathcal R_{\slashed E}$ in which the motional electric field vanishes. The oscillations have disappeared and $\gamma'_{\rm obs}(t)$ evolves as the particle gains or loses energy through Fermi processes. Finally, the dashed red line shows the reconstructed particle history $\gamma'_{\rm th}(t)$, using Eq.~(\ref{eq:Rco-evol}) with initial condition  $\gamma'_{\rm th}(t_0)=\gamma'_{\rm obs}(t_0)$ at the initial time $t_0=1.8\, \ell_{\rm c}/c$.

In the upper panel, we observe that the match between the reconstructed and the observed trajectories is rather tight in regions where the frequency of the oscillations increase, {\it e.g.} $3\,\ell_{\rm c}/c\lesssim t\lesssim 4 \ell_{\rm c}/c$. This is not unexpected, insofar as an increase in the frequency of oscillations corresponds to a decrease in the particle gyroradius, and the model works to order $r_{\rm g}/\ell_{\rm c}$. On the contrary, at later times $t\gtrsim 5\ell_{\rm c}/c$, the particle has achieved a larger energy, and it seemingly propagates in a region of lower-than-average magnetic strength, hence the ratio $r_{\rm g}/\ell_{\rm c}$ is no longer small compared to unity, as evidenced by the time scale of the oscillations. Deviations from the observed trajectory can thus be expected at that stage, although they remain rather mild. 

In the lower panel, the energy history is well reconstructed at early times $t\lesssim 3\ell_{\rm c}/c$. We observe an offset in the vertical direction between the predicted and observed trajectories at later times, although those two histories maintain a rather strong degree of correlation. Had we chosen as initial time $t_0\simeq 3-3.5\,\ell_{\rm c}/c$, we would thus have obtained a nice match to the observed history at late times. In effect, the departure between the model and the simulation is limited to the interval $\sim 2.7-3.3\,\ell_{\rm c}/c$, and likely related to some small scale effect. As mentioned before, this observation has motivated our choice to adopt two timescales for the comparison of the model to the simulations: one reduced timescale of the order of $1\,\ell_{\rm c}/c$, and one integrating over the whole history.

\subsection{3D forced turbulence MHD simulation}

Finally, we compare the model to trajectories of test particles that were tracked in the 3D forced MHD simulation of the JHU turbulence database~\cite{2008JTurb...9...31L,2013Natur.497..466E}. This 3D direct numerical simulation solves the incompressible MHD equations on a $1\,024^3$ periodic grid with a time resolution $\delta t \simeq 0.04\,\delta x$. The database output provides $1024$ time snapshots, with sampling interval $10\,\delta t$. The simulation is visco-resistive, with magnetic Prandtl number unity, and magnetic Reynolds number $R_\lambda \sim 140$ at the Taylor scale $\lambda \sim 1.05\times 10^{-2}\,L$; here, $L$ represents the size of one side of the simulation cube and the Taylor scale is defined as $\lambda=\left(5\, \int {\rm d}k\, S_k/\int {\rm d} k\, k^2 S_k\right)^{1/2}$, where $S_k$ denotes the one-dimensional power spectrum of magnetic fluctuations. The Alfv\'en velocity is $v_{\rm A}=0.41c$ and the rms velocity $\langle\delta u^2\rangle^{1/2}\simeq 0.4c$. The turbulence is excited through an external force acting on the velocity field at a stirring wavenumber $k_{\rm f} \simeq 12.6\, L^{-1}$. At the reference time $t=0$, the simulation, as made available on the database, has already achieved a steady state.

\begin{figure}
\includegraphics[width=0.48\textwidth]{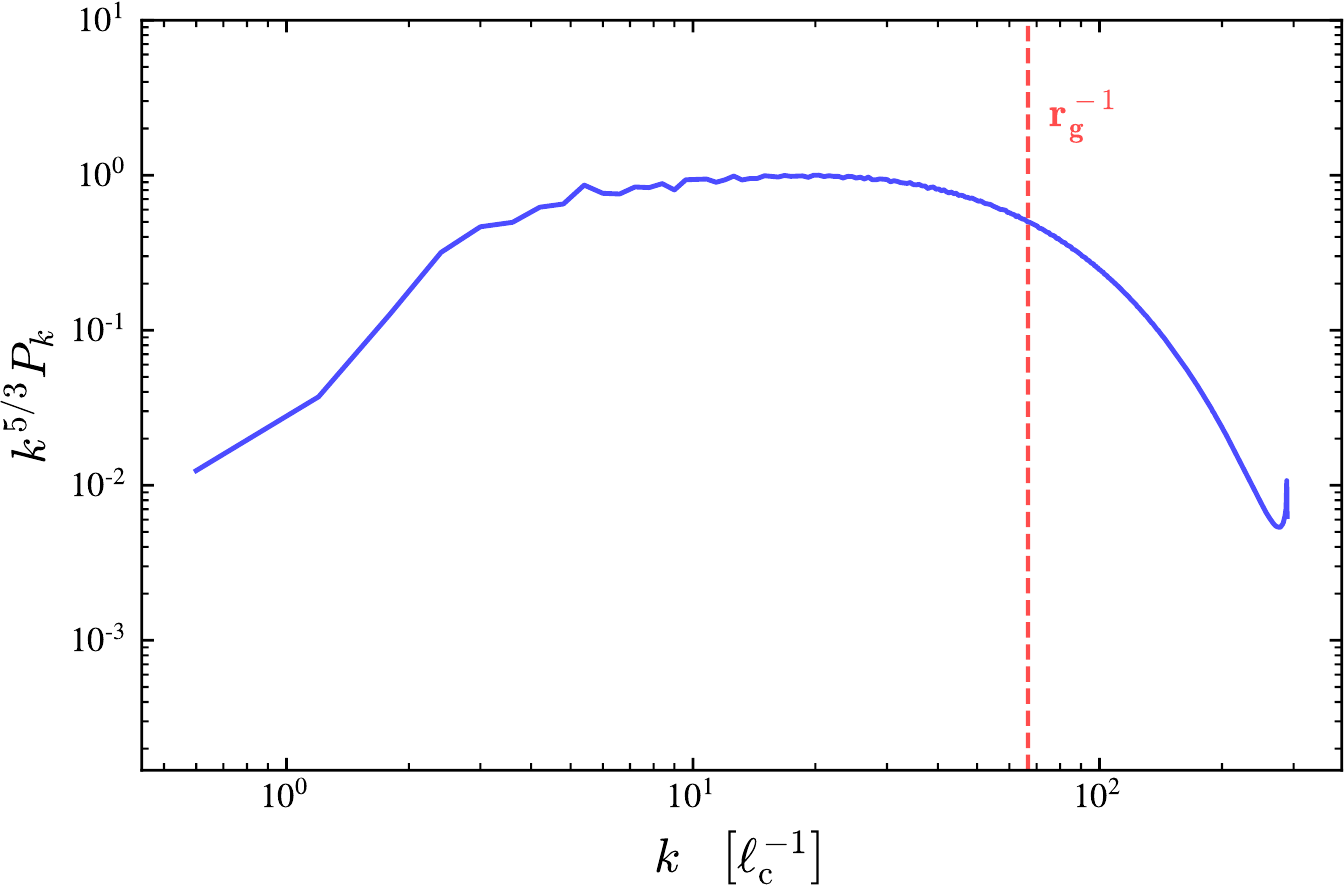}
 \caption{Magnetic power spectrum of the 3D MHD simulation, rescaled by $k^{5/3}$, vs wavenumbers in units $\ell_{\rm c}^{-1}$. The inverse gyroradius of the test particles is indicated by a dashed red line. It lies at the transition between the inertial and the dissipative range.
 \label{fig:pw-MHD} }
\end{figure}

The integral scale of the turbulence, as defined in the database, is $L_w\sim 0.1$ in units of the cube size. The simulation volume thus comprises many coherence cells of the turbulence, hence the effective dynamic range is restricted to $L_w/\delta x \sim 100$. The power spectrum of magnetic fluctuations is shown in Fig.~\ref{fig:pw-MHD}. It correspondingly reveals a lack of power at wavenumbers $k\lesssim \,3\,\ell_{\rm c}^{-1}$ followed by the $k^{-5/3}$ scaling in the inertial range. We adopt here $\ell_{\rm c}=0.1\,L$.  

\begin{figure}
\includegraphics[width=0.48\textwidth]{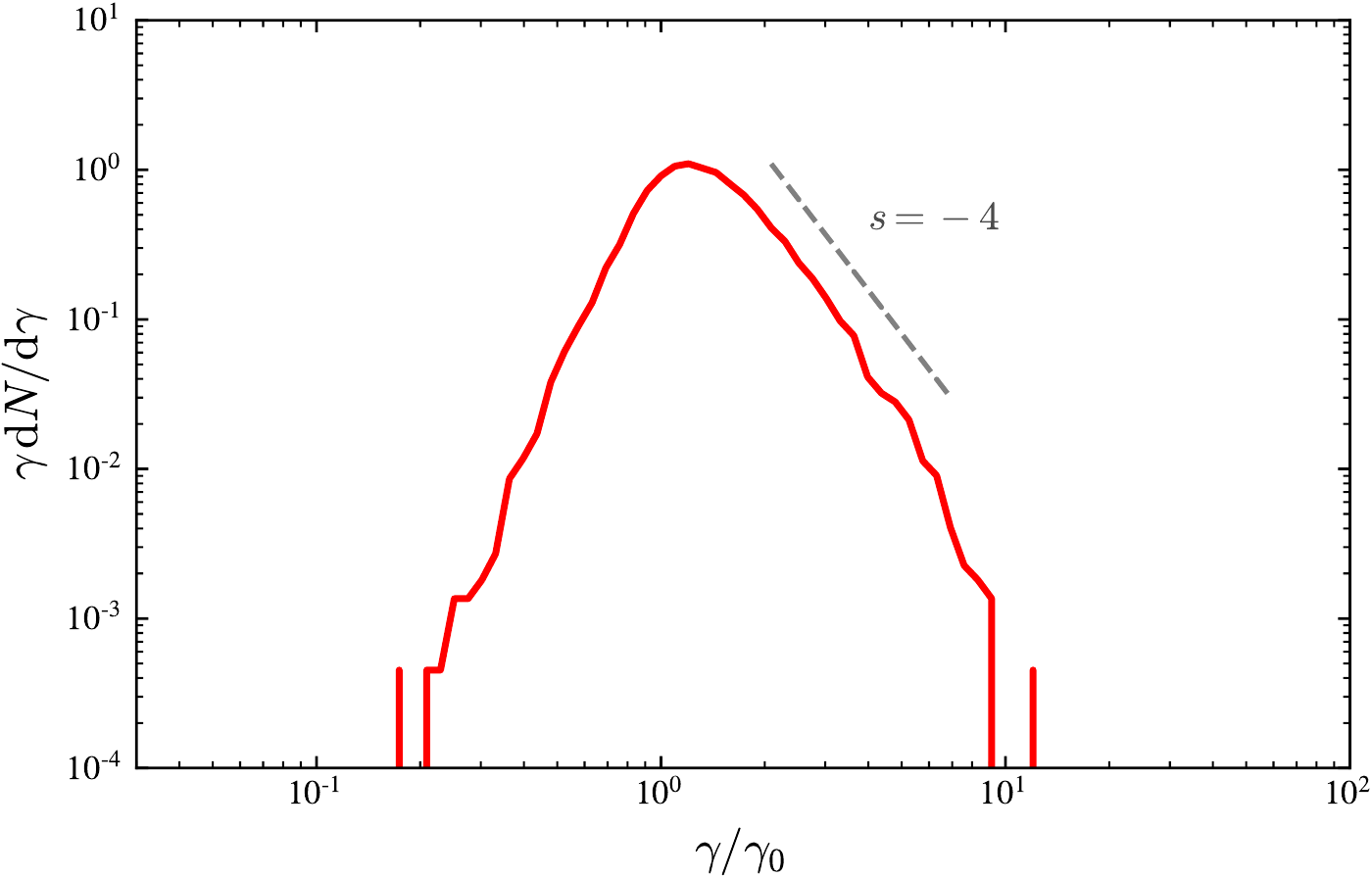}
 \caption{Energy distribution of test particles propagated in the 3D MHD simulation, at a time $t\simeq 4\,\ell_{\rm c}/c$. At the initial time $t=0$, all particles shared a common Lorentz factor $\gamma_0$, corresponding to a gyroradius $r_{\rm g}\simeq 0.02\,\ell_{\rm c}$. The spectrum takes a power law shape at large energies, with spectral index $s\simeq-4$, assuming ${\rm d} N/{\rm d}\gamma\propto \gamma^s$.
 \label{fig:sp-MHD} }
\end{figure}

We follow test particles with a gyroradius $r_{\rm g} \simeq 2 \delta x$, in order to maintain $r_{\rm g}/\ell_{\rm c}$ as small as possible while preserving a reasonable reconstruction of the particle trajectory. As can be seen from Fig.~\ref{fig:pw-MHD}, the inverse gyroscale $r_{\rm g}^{-1}$ lies at the transition between the inertial and the dissipative range. The effective rigidity is $2\pi r_{\rm g}/\ell_{\rm c}\sim 0.1$. Experiments conducted with a gyroradius twice as large provide similar results. We propagate $24\,000$ particles over $4.2\,\ell_{\rm c}/c\sim 200\,r_{\rm g}$; those particles were initialized with a common Lorentz factor (in the simulation frame), corresponding to the desired gyroradius, at random positions and velocity orientations. For each test particle, we integrate its trajectory over the duration of the simulation, using a numerical Monte Carlo code which, at each time step, queries the database to retrieve the values of the magnetic and velocity field at the particle location. The field values are determined at the particle spatial location using high-order ($4^{\rm th}$ or $6^{\rm th}$) Lagrangian interpolation. Although we sample the particle trajectory with a time step of $0.1\,r_{\rm g}/c$, we do not seek to interpolate the field values at the corresponding intermediate times, and rather use the values calculated from the nearest snapshot. Given that the typical velocity on the grid size is of the order of $\left(\delta x/\ell_{\rm c}\right)^{1/3}\,\langle\delta u^2\rangle^{1/2}\sim 0.050\,c$ -- assuming a standard Kolmogorov scaling -- this represents a reasonable approximation. This also allows us to maintain the computational time within reasonable limits, since computational time is here dominated by the queries to the database, which are performed online.

The code computes the electric field at the particle location using ideal Ohm's law, then advances the particle using a Boris pusher. All along the trajectories, we record the time and space derivatives of the magnetic and electric fields; the latter is computed from the magnetic and velocity derivatives. Those derivatives, provided by the database as 4$^{\rm th}$-order centered finite differencing, are used to calculate the quantities that enter the force terms in Eq.~(\ref{eq:Rco-evol}), as for the PIC simulation. The database does not directly provide time derivatives; those are thus calculated using first-order finite differencing from values obtained at consecutive times. We note that the force terms that enter Eq.~(\ref{eq:Rco-evol}) are dominated by the spatial derivatives in the sub- or mildly relativistic conditions of the present MHD simulation. 

In Fig.~\ref{fig:sp-MHD}, we plot the resulting energy distribution after a time $t\simeq 4\,\ell_{\rm c}/c$. It reveals a powerlaw tail at large momenta, as in the PIC simulation. To our knowledge, such a behavior had not been observed in time-evolving MHD simulations before. The spectral index $s\simeq -4$ is somewhat larger (in absolute value) than that observed in the PIC simulation, as expected for particle acceleration in a turbulence of smaller magnetization level~\cite{18Comisso,2021PhRvD.104f3020L}.

\begin{figure}
\includegraphics[width=0.48\textwidth]{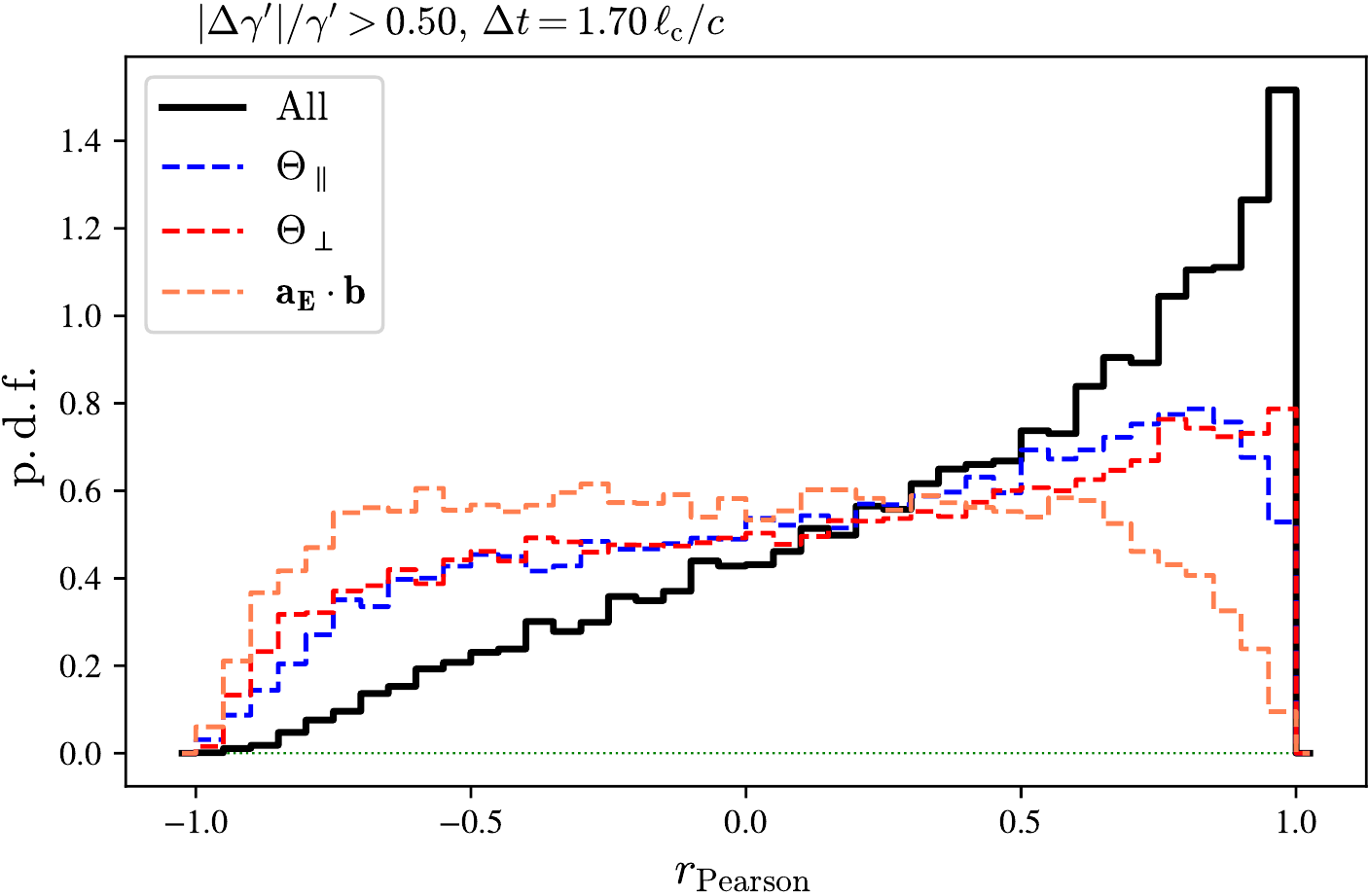}
 \caption{Histogram of correlation coefficients between the expected and observed evolution of $\gamma'$ along blocks of the energy histories of test particles that have been propagated in the 3D MHD simulation. The initial Lorentz factor for all particles is $\gamma(t=0)=10$ (simulation frame).
 \label{fig:hist-ck-MHD} }
\end{figure}

\begin{figure}
\includegraphics[width=0.48\textwidth]{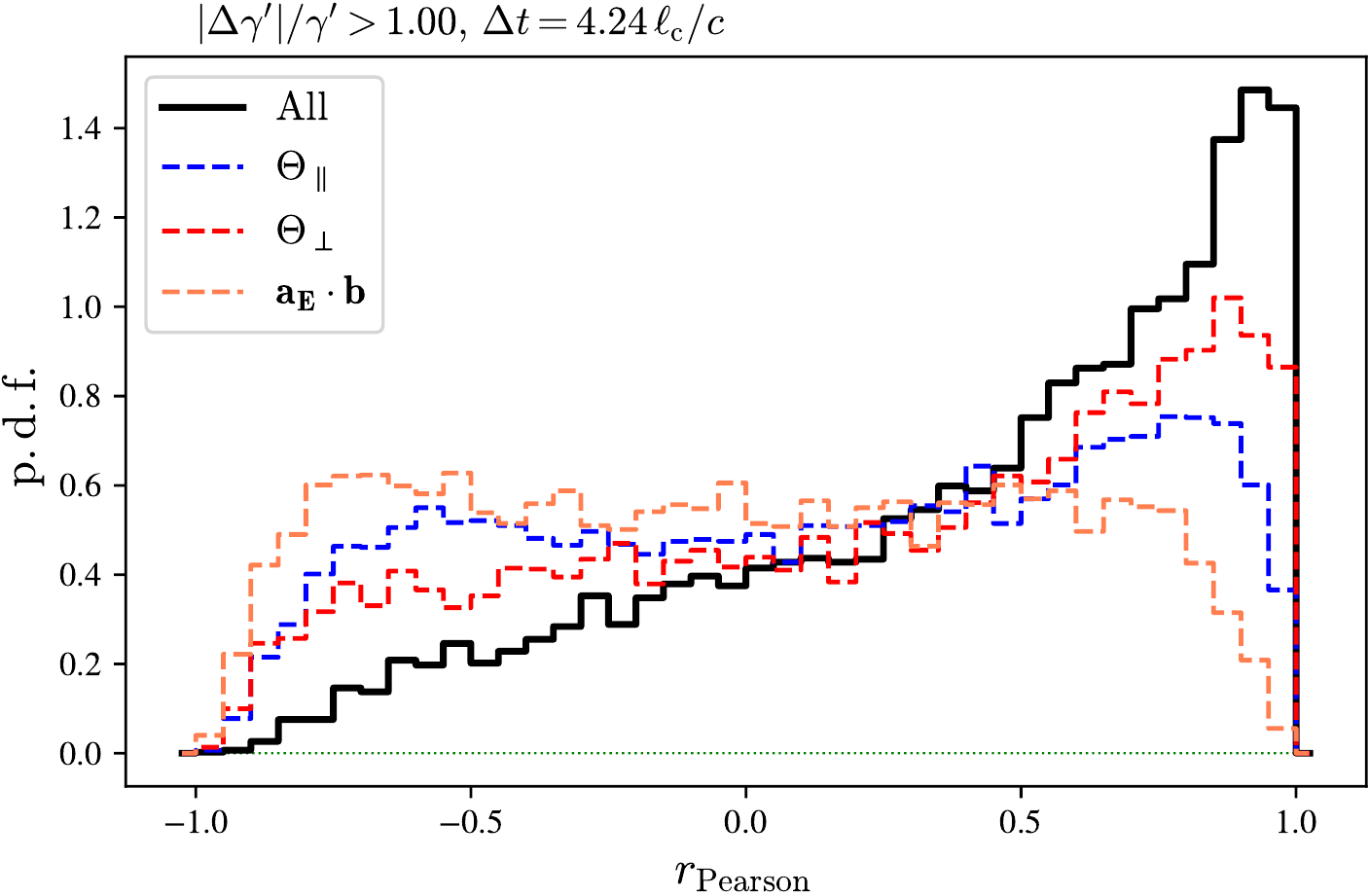}
 \caption{Same as Fig.~\ref{fig:hist-all-MHD} (3D MHD simulation), now considering the whole trajectory of each test particle. 
 \label{fig:hist-all-MHD} }
\end{figure}

In Figs.~\ref{fig:hist-ck-MHD} and \ref{fig:hist-all-MHD}, we present the histograms of the correlation coefficients.
As threshold of energy variation and duration of integration, we have adopted $\vert\Delta\gamma'\vert/\gamma'>0.5$ and $\Delta t \simeq 1.7\,\ell_{\rm c}/c$ in a first case (Fig.~\ref{fig:hist-ck-MHD}), $\vert\Delta\gamma'\vert/\gamma'>1$ and $\Delta t\simeq 4.2\,\ell_{\rm c}$ in a second one (Fig.~\ref{fig:hist-all-MHD}). The lesser threshold in energy variation and longer duration of the interval, comparatively to the PIC simulations, are meant to compensate for slower acceleration in the present simulation.

We recover here a high degree of correlation, as observed in the 3D PIC simulation. Interestingly, that degree of correlation is, in the present case, substantially higher when all force terms are combined together using Eq.~(\ref{eq:Rco-evol}) to reconstruct $\gamma'_{\rm th}(t)$, than when they are taken individually. We also note that both $\Theta_\parallel$ and $\Theta_\perp$ seem to provide contributions with a net positive degree of correlation, when taken individually, while the influence of $\boldsymbol{a_E}\cdot\boldsymbol{b}$ is not visible here. Interestingly, the degrees of correlation of $\Theta_\perp$ and $\Theta_\parallel$ appear to be of the same or order of magnitude, while the PIC simulations showed a clear dominance of $\Theta_\parallel$. We cannot identify here the exact reason why that is so, but we speculate that this difference may indicate that their relative contribution depends on how the turbulence is driven: incompressible turbulence driven by external velocity fluctuations in the MHD case {\it vs} compressible turbulence driven by magnetic perturbations in the kinetic regime. Recent PIC simulations have similarly demonstrated that the efficiency of acceleration depends on the stirring procedure~\cite{2021ApJ...922..172Z}. This difference may also be affected by the different velocity regimes (sub-relativistic for MHD, relativistic for PIC).

With respect to the test of our model, we stress here the significance of observing such a significant degree of correlation for both 3D PIC and MHD simulations, up to the above difference in individual contributions: on the ``large'' length scales that we are interested in (comparatively to the kinetic scales), both should in principle reproduce the same physics of acceleration; however, both rely on different schemes of approximations. In particular, the MHD case neglects all kinetic effects and all deviations of Ohm's law that are inherently included in PIC simulations.

\section{Summary -- conclusions}\label{sec:conc}
In the present work, we report on a statistical comparison of a recent model of non-resonant particle acceleration in magnetized turbulence {\it vs} 2D and 3D kinetic numerical simulations as well as a 3D (incompressible) MHD simulation. This model describes energization as the continuous interaction of the particle with the random velocity flow of the turbulence, in the frame of ideal MHD; it can thus be regarded as the direct generalization  to a continuous turbulent flow of the original Fermi picture of discrete, point-like interactions~\cite{2019PhRvD..99h3006L,2021PhRvD.104f3020L}. It does so by following the evolution of the particle momentum in the frame $\mathcal R_{\slashed E}$ that moves with the magnetic field lines at velocity $\boldsymbol{v_E}=\boldsymbol{E}\times \boldsymbol{B}/B^2$, and where the electric field vanishes.  This allows to relate the sources of energy gains and losses to the gradients of the velocity field $\boldsymbol{v_E}$, and more particularly to three main contributions: an inertial term $\boldsymbol{a_E}\cdot\boldsymbol{b}$, a longitudinal shear term $\Theta_\parallel$ and a perpendicular compressive mode $\Theta_\perp$, the notions of longitudinal/perpendicular being defined relative to the mean magnetic field direction at that location. To lowest order in the ratio of particle gyroradius to coherence scale of the turbulence, $r_{\rm g}/\ell_{\rm c}$, the evolution of the particle energy is captured by Eq.~(\ref{eq:Rco-evol}).

To test this theoretical model, we have conducted PIC simulations of 2D decaying turbulence, of 2D and 3D driven turbulence in the relativistic regime $v_{\rm A}\sim c$, and we have made use of the 3D forced MHD simulation of the JHU-database. We have then followed the time histories of the energy for a large sample of particles and compared the observed time histories to those reconstructed by the model. For what regards the MHD simulation, we have propagated test particles through the simulation, properly taking into account the time evolution of the fields. In all simulations, we have selected particles whose inverse gyroradius corresponds to wavenumbers at or below the transition between the inertial and the dissipative range of the turbulence, in order to test the model in conditions in which it applies, namely a ratio $r_{\rm g}/\ell_{\rm c}$ as small as possible and near-MHD conditions. To obtain the reconstructed particle histories, we have extracted from the simulations the quantities $\boldsymbol{a_E}\cdot\boldsymbol{b}$, $\Theta_\parallel$ and $\Theta_\perp$ then used Eq.~(\ref{eq:Rco-evol}) at each point along the particle trajectory to integrate in time the particle energy using Eq.~(\ref{eq:Rco-evol}). We have then computed for each particle trajectory a Pearson correlation test between the two histories (observed vs reconstructed), then derived from the sample of particles a probability density of the correlation coefficients $r$. A perfect adequation of the model to the date would translate in a probability density sharply peaked around $+1$, while an complete inadequacy would rather yield a featureless, roughly uniform histogram over the interval $[-1,\,+1]$. We have verified this using Monte Carlo simulations of test-particle transport in a synthetic turbulence composed of a sum of linear eigenmodes of the plasma, see App.~\ref{app:synthetic}.

Our main result is that we observe a clear-cut correlation between the model predictions and the numerical experiments, with histograms of the Pearson correlation coefficients distinctly peaked around $+1$, for all numerical simulations. This indicates that the non-resonant model can successfully account for the bulk of particle energization through stochastic Fermi processes. Let us recall here that particle acceleration in a magnetized turbulence appears to proceed in two distinct stages: an injection into the non-thermal population through non-ideal electric fields, then acceleration \`a la Fermi up to much higher energies~\cite{18Comisso,2019ApJ...886..122C}. The model and the test that we study here thus apply to the second stage, where the influence of non-ideal electric fields can be neglected.

In our PIC numerical simulations, we observe that the longitudinal shear term $\Theta_\parallel$ appears to provide the dominant contribution to particle energization, because the correlation histogram when neglecting the other two force terms in the theoretical reconstruction of the energy histories lies close to that obtained when considering all force terms. This longitudinal shear term can be depicted as a form of slingshot acceleration in a moving, curved magnetic field, as in the Fermi type-B interaction of the original Fermi model~\cite{49Fermi}. Contrariwise, the MHD simulation reveals about similar degrees of correlation of $\Theta_\perp$ and $\Theta_\parallel$, with a slight preference for the former, which characterizes magnetic mirroring effects, or type-A Fermi interactions. This MHD simulation also shows a significantly higher degree of correlation when all contributions are summed together as in the model Eq.~(\ref{eq:Rco-evol}) than when only one force term is considered individually, and the other two discarded.

This difference in contributions between the MHD and the PIC simulations suggests that the physics of acceleration, in particular the dominant energization process, depends on the stirring process, on the nature of the turbulence and/or the velocity regime: while the (sub-relativistic) turbulence of the MHD simulation is by construction incompressible and forced through solenoidal velocity motions, the (relativistic) turbulence in the PIC simulations is compressible and driven through external magnetic perturbations. A dependence of the energy distribution of accelerated particles on the stirring process (solenoidal vs compressible) has been noted before in Ref.~\cite{2021ApJ...922..172Z}.

More work is clearly needed to shed light on those issues and to develop further the above non-resonant model. It would be useful, in particular, to extend the microphysical random walk introduced in~\cite{2021PhRvD.104f3020L} in extract predictions for the particle energy spectrum, and to derive a kinetic equation for the distribution function. More work is also needed to understand how the present picture extends into the sub-relativistic regime, to other domains of the turbulence cascade ({\it e.g.}, dissipative) and to electron-ion plasmas. This will be the subject of future work.

\begin{acknowledgments}
The possibility to use the resources of the JH Turbulence Database (JHTDB), which is developed as an open resource by the Johns Hopkins University, under the sponsorship of the National Science Foundation, as well as the tools made available for public use, is gratefully acknowledged. This work has been supported in part by the Sorbonne Universit\'e DIWINE Emergence-2019 program, by the ANR (UnRIP project, Grant No.~ANR-20-CE30-0030) and by the Centre National d'\'Etudes Spatiales (CNES). We acknowledge GENCI-TGCC for granting us access to the supercomputer IRENE under Grants No.~2019-A0050407666, 2020-A0080411422 and 2021-A0080411422 to run PIC simulations. L.S. acknowledges support from the Cottrell Scholars Award, NASA 80NSSC20K1556, NSF PHY-1903412, DoE DE-SC0021254 and NSF AST-2108201. C.D. acknowledges support from NSF grant AST 1903335 and NASA grant NNX17AK55G.
\end{acknowledgments}
\bigskip

\appendix

\section{Test on synthetic turbulence}\label{app:synthetic}
In this Appendix, we conduct comparisons similar to those discussed in the main text for PIC and MHD simulations, although using a synthetic turbulence constructed as a sum of non-interacting linear MHD eigenmodes. Quasilinear theory predicts that particle energization takes place through two types of interactions: gyroresonant interactions of the form $k_\parallel v_\parallel - \omega \simeq n\,c/r_{\rm g}$ ($n\in \mathbb{Z^\star}$) and Landau-synchrotron resonances $k_\parallel v_\parallel - \omega = 0$. Gyroresonant interactions can take place for all modes, at least in the absence of local anisotropy \`a la Goldreich-Sridhar~\cite{2000PhRvL..85.4656C,2002PhRvL..89B1102Y}, while the Landau-synchrotron (transit-time damping) are specific to fast magnetosonic modes. We recall that those transit-time damping interactions are related to magnetic mirroring effects, which are captured by the theoretical model of non-resonant interactions ($\Theta_\perp$ contribution). However, that model does not predict any gyroresonant interaction. Conversely, Alfv\'en modes predict, in the linear limit, $\Theta_\parallel=0$ as well as $\Theta_\perp=0$, while fast magnetosonic modes lead to $\Theta_\parallel=0$ but $\Theta_\perp \neq 0$~\cite{2021PhRvD.104f3020L}.

The numerical code used to build the synthetic turbulence and track particles therein is presented in \cite{2020PhRvD.102b3003D}. In brief, particle trajectories are integrated using a Bulirsch-Stoer algorithm. At each timestep, the electromagnetic and velocity fields at the location of particles are constructed as the sum of a background magnetic field and the superposition of the fluctuations carried by a collection of waves with dispersion relation and polarizations of (special relativistic) MHD eigenmodes. The electric field is derived from the total magnetic field and total velocity field through ideal Ohm's law. The wavevectors and amplitudes of the waves are initialized so as to achieve the desired power spectrum of turbulence over a range of scales $[L_\text{min}, L_\text{max}]$. Particles are injected along random directions in different turbulence realizations with the energy corresponding to the gyroradius of interest. To reconstruct the energy histories using Eq.~(\ref{eq:Rco-evol}), we calculate the spatial and temporal derivatives of the magnetic field and the velocity field, then derive those of the electric field through ideal Ohm's law. In this synthetic turbulence, the derivatives can be expressed analytically in terms of  the plane wave expansion.

We conduct two experiments on such synthetic turbulence comprised of 256 modes, with wavelengths extending from $L_\text{max}=\ell_{\rm c}$ down to $L_\text{min}=L_\text{max}/100$. In experiment (A), we simulate a turbulence of isotropic fast magnetosonic modes with $\delta B/B_0=1$, Alfv\'en velocity $v_\text{A}=0.6\,c$, sound velocity $v_{\rm s}\ll v_{\rm A}$, which implies a phase velocity for each wave $v_{\rm F}\simeq v_{\rm A}$. Isotropic means here that the turbulent magnetic power spectrum does not depend on the direction of the wavenumber; its scaling is assumed to follow Kolmogorov $S_k\propto k^{-5/3}$. We inject particles with a gyroradius outside the range of scales of the turbulence, $r_\text{g} = 0.1\,L_\text{min}$. In that configuration, gyroresonant interactions are suppressed because restricted to high harmonics (large $n$) so that acceleration is dominated by transit time damping acceleration~\cite{19Teraki}. We thus expect the theoretical model to provide a fair reconstruction of the trajectories, at least its $\Theta_\perp$ part. In a second experiment, (B), we simulate an opposite situation, namely a turbulence of Alfv\'en modes with $v_\text{A}=0.6\,c$, $\delta B/B_0=1$ and set the gyroradius of the particles to fall in the range of wavelengths of the turbulence, $r_{\rm g}=0.1\,L_{\rm max}=10\,L_{\rm min}$, which thus permits gyroresonant interactions at the first harmonic $n=1$. We simulate here simple Alfv\'en waves, meaning that we neglect any wave damping term and that we assume an isotropic Kolmogorov cascade. Our aim indeed is to bring to light the effect of gyroresonances, or rather, the lack of correlation between observed and reconstructed trajectories in a situation in which most of energy gain is known to result from gyroresonant interactions; we thus deliberately render those resonances sharp. We choose Alfv\'en waves in order to erase any magnetic mirroring effect. Consequently, we expect the theoretical model to behave poorly in that limit, given that it ignores such gyroresonances, by construction.

\begin{figure}
\includegraphics[width=0.48\textwidth]{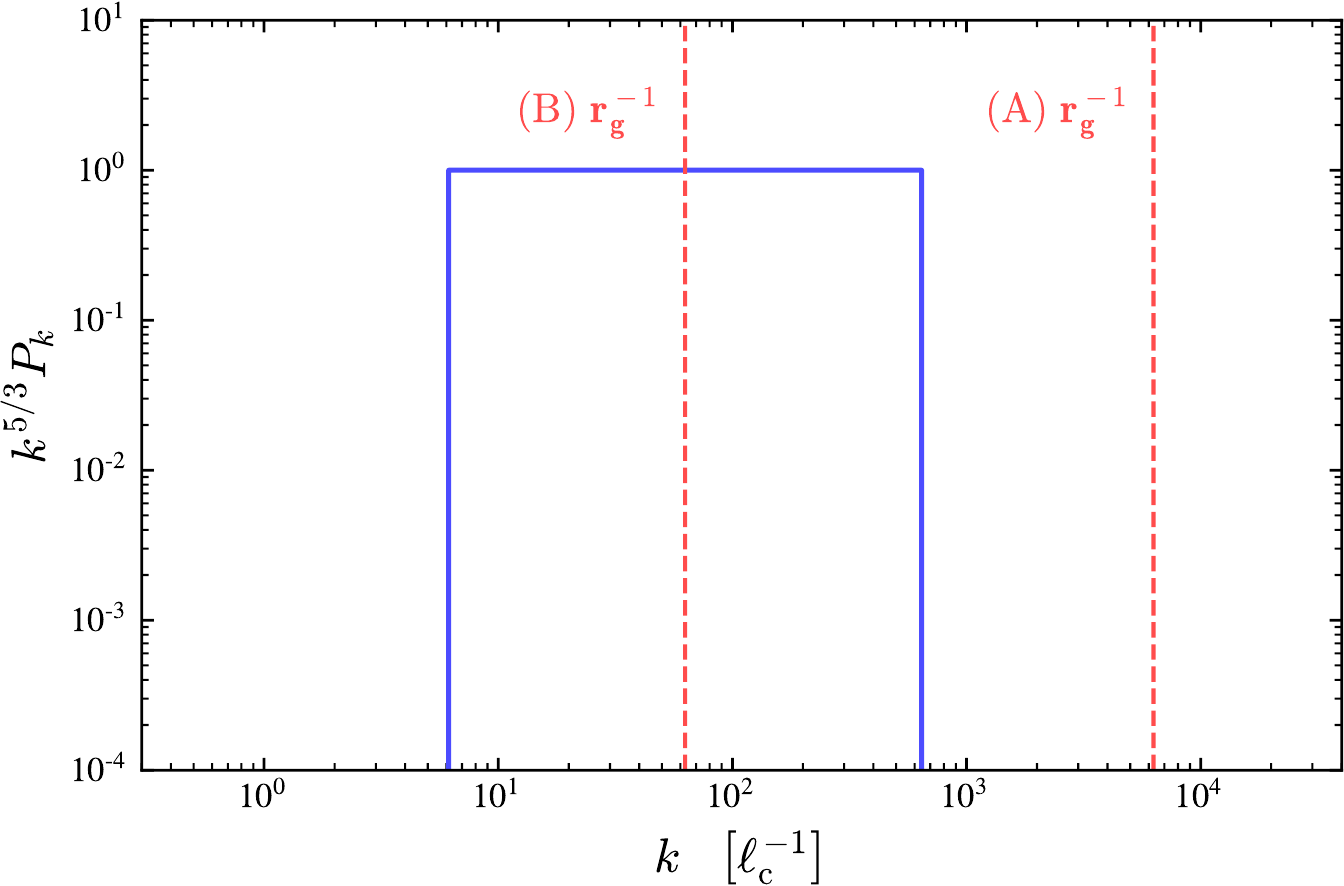}
 \caption{Power spectrum of synthetic turbulence, with locations of the inverse gyroradii indicated by dashed lines for the two experiments (A), corresponding to fast mode turbulence, and (B), for Alv\'en modes.
 \label{fig:spec-app1} }
\end{figure}

Figure~\ref{fig:spec-app1} shows the power spectrum (normalized by $k^{5/3}$) of magnetic fluctuations in this synthetic wave turbulence, with the locations of $r_{\rm g}^{-1}$ indicated as dashed lines for both models: model (A) with $r_{\rm g}$ below the minimum scale, and model (B) with $r_{\rm g}$ in the inertial range.

\begin{figure}
\includegraphics[width=0.48\textwidth]{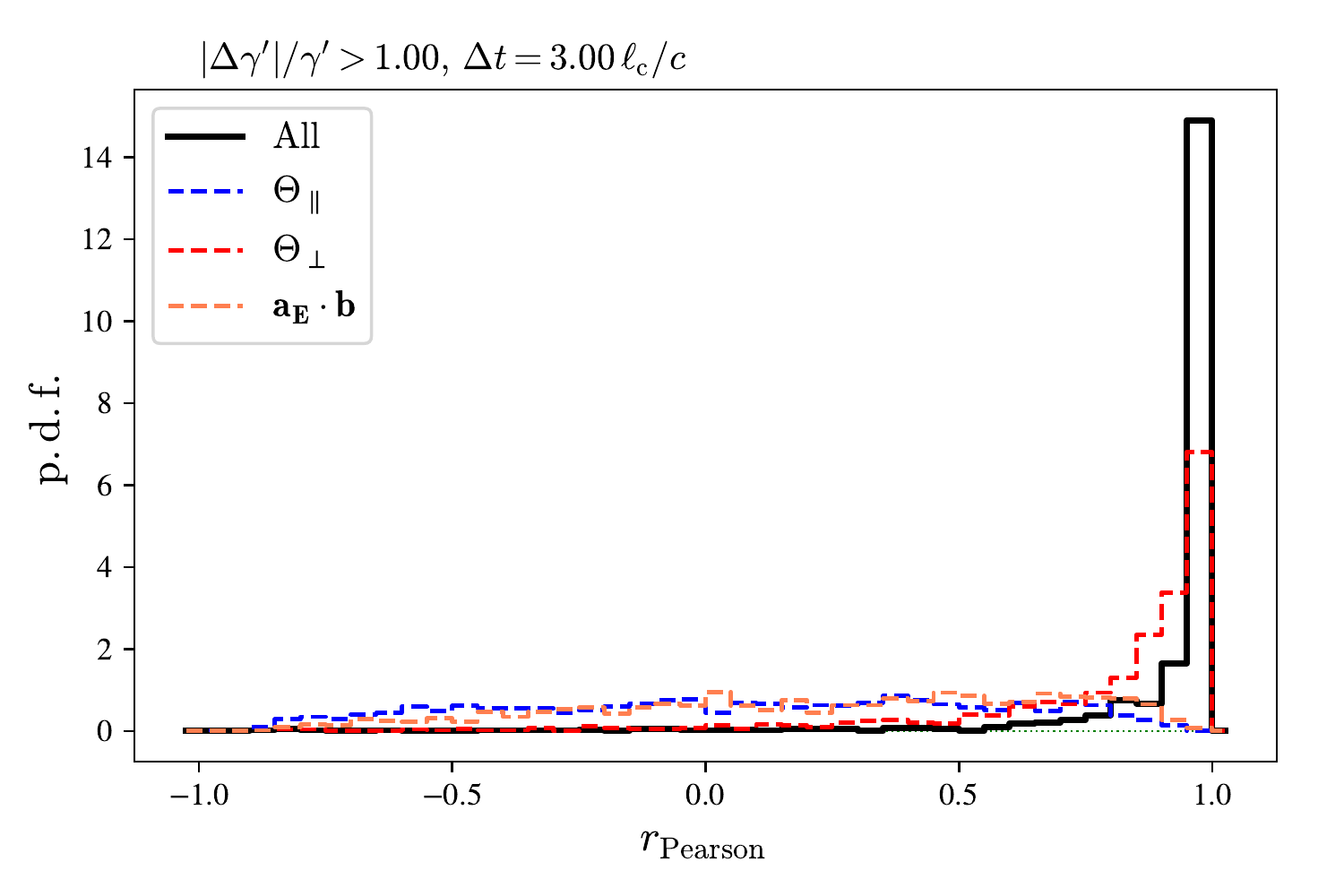}
 \caption{Probability density of correlation coefficients in experiment (A), in which particle acceleration takes place through interactions with magnetic mirror modes (transit-time damping in fast mode synthetic turbulence). The model fully captures the energy gains, with a strongly dominant contribution of $\Theta_\perp$, as expected. 
 \label{fig:hist-app1} }
\end{figure}

The histogram of the probability density of the Pearson correlation coefficients between the observed and reconstructed trajectories for model (A) is shown in Fig.~\ref{fig:hist-app1}. The concentration of the probability density of $r$ around $+1$ indicates that, as anticipated, the model is highly successful in reproducing the trajectories, most notably so for the magnetic mirror ($\Theta_\perp$) contribution. A closer inspection reveals that the non-resonant model, summing over the contributions of all three force terms, provides a better match to the observed energy histories than the contribution of magnetic mirrors alone, as its probability density is more sharply peaked around $+1$. The difference comes from the inertial term $\boldsymbol{a_E}\cdot\boldsymbol{b}$, which provides a net contribution, as evidenced by its overall positive degree of correlation. Within the frame of the model that we are testing, this is not altogether surprising, insofar as this inertial term characterizes the influence of accelerations/decelerations of the frame $\mathcal R_{\slashed E}$ in which the notion of a magnetic mirror can be properly defined. In that sense, the inertial term should not be left aside when considering the influence of  $\Theta_\perp$ (or $\Theta_\parallel$, for similar reasons).

\begin{figure}
\includegraphics[width=0.48\textwidth]{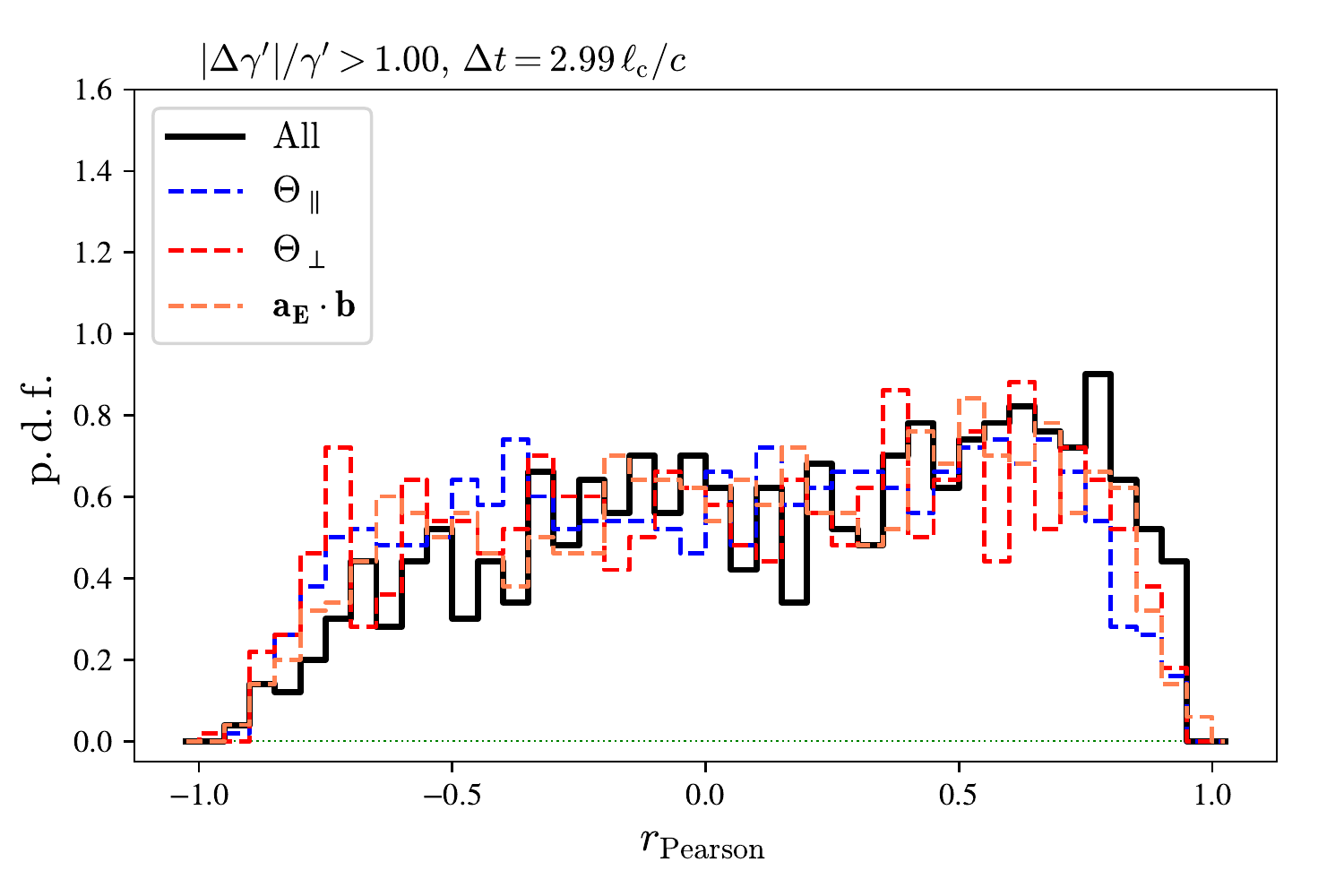}
 \caption{Same as Fig.~\ref{fig:hist-app2}, for case (B), in which particle acceleration now takes place through gyroresonant interactions. In that case, the model is unable to reproduce the energy gains, hence the pdf of correlation coefficients appears devoid of structure, revealing no particular preference for values close to $+1$.
 \label{fig:hist-app2} }
\end{figure}

Figure~\ref{fig:hist-app2} shows the corresponding histograms in the case of model (B), which appear relatively structureless and uniformly distributed over the interval $[-1,\,+1]$. There appears to be a slight bias toward positive values of the correlation coefficients, but nothing of the sort discussed previously in Sec.~\ref{sec:num} for PIC or MHD simulations. This suggests that most of the energy gains/losses are indeed not captured by the non-resonant model in the present case and that gyroresonant wave-particle interactions do not leave a strong signature in this histogram of correlation coefficients. The positive correlations observed in Sec.~\ref{sec:num} can thus be interpreted as genuine evidence in favor of non-resonant acceleration.

\bibliographystyle{apsrev4-1}

\bibliography{refs}

\end{document}